\documentclass[12pt,dvipsnames]{article}
\usepackage[top=80pt,bottom=85pt,left=85pt,right=85pt]{geometry}

\pdfoutput=1
\usepackage[utf8]{inputenc}
\usepackage[T1]{fontenc} 
\usepackage[english]{babel} 
\usepackage{braket}
\usepackage{textcomp}
\usepackage[dvipsnames]{xcolor} %Package for fancy coloros
\usepackage{physics}
\usepackage{tabularx} 
\usepackage{ragged2e}
\usepackage{booktabs} 
\usepackage{rotating}
\usepackage{makecell,booktabs}
\usepackage{multirow} 
\usepackage{comment} 
\usepackage{amsmath}
\usepackage{cite}
\usepackage[all]{xy}

\usepackage{tabularray}

\usepackage{cite}

\usepackage{tikz}
\usetikzlibrary{math} 
\usetikzlibrary{3d} 
\usepackage{tikz-3dplot}
\usepackage{tikz-cd} 
\usepackage{microtype}
\usepackage{amsthm}
\usepackage{mathrsfs} 
\usepackage[strict]{changepage}

\tikzcdset{
  cells={font=\everymath\expandafter{\the\everymath\displaystyle}},
}

\newcommand{\xdownarrow}[1]{%
  {\left\downarrow\vbox to #1{}\right.\kern-\nulldelimiterspace}
}

\usepackage{amssymb} 
\usepackage{bbold}
\usepackage{subcaption} 
\DeclareCaptionFormat{custom}

\usepackage[pdftex,colorlinks=true]{hyperref} 
\hypersetup{urlcolor=RedOrange, citecolor=burntorange, linkcolor=OliveGreen}

\usepackage{float} 
\usepackage{todonotes}

\usepackage{cleveref}

\newcolumntype{E}{>{\hfil$}p{0.65cm}<{$\hfil}}

\newcolumntype{L}{>{\hfil$}p{16cm}<{$\hfil}}

\newcolumntype{D}{>{\hfil$}p{7.4cm}<{$\hfil}}
\newcolumntype{C}{>{\hfil$}p{3cm}<{$\hfil}}
\newcolumntype{P}{>{\hfil$}p{7.7cm}<{$\hfil}}
\newcolumntype{F}{>{\hfil$}p{5.7cm}<{$\hfil}}
\newcolumntype{S}{>{\hfil$}p{1.8cm}<{$\hfil}}
\newcolumntype{R}{>{\hfil$}p{5.2cm}<{$\hfil}}
\newcolumntype{U}{>{\hfil$}p{4.2cm}<{$\hfil}}
\newcolumntype{Q}{>{\hfil$}p{6.4cm}<{$\hfil}}
\newcolumntype{T}{>{\hfil$}p{1.9cm}<{$\hfil}}
\newcolumntype{V}{>{\hfil$}p{5.8cm}<{$\hfil}}
\newcolumntype{H}{>{\hfil$}p{1.8cm}<{$\hfil}}
\newcolumntype{A}{>{\hfil$}p{6cm}<{$\hfil}}
\newcolumntype{B}{>{\hfil$}p{2cm}<{$\hfil}}
\makeatletter
\newcommand\xleftrightarrow[2][]{%
  \ext@arrow 9999{\longleftrightarrowfill@}{#1}{#2}}
\newcommand\longleftrightarrowfill@{%
  \arrowfill@\leftarrow\relbar\rightarrow}
\makeatother

\renewcommand{\vec}[1]{\mathbf{#1}}

\numberwithin{equation}{section}

\definecolor{burntorange}{rgb}{0.8, 0.33, 0.0}
\definecolor{cambridgeblue}{rgb}{0.64, 0.76, 0.68}
\definecolor{caribbeangreen}{rgb}{0.0, 0.8, 0.6}
\definecolor{celadon}{rgb}{0.67, 0.88, 0.69}
\definecolor{champagne}{rgb}{0.97, 0.91, 0.81}
\definecolor{cream}{rgb}{1.0, 0.99, 0.82}
\definecolor{cyan(process)}{rgb}{0.0, 0.72, 0.92}
\definecolor{brilliantlavender}{rgb}{0.96, 0.73, 1.0}
\definecolor{candypink}{rgb}{0.89, 0.44, 0.48}

\begin{document}

\begin{titlepage}

\phantom{wowiezowie}

\vspace{-1cm}

\begin{center}

{\Huge {\bf Symmetries Beyond Branes:}}

\vspace{0.2cm}

{\Huge {\bf Geometric Engineering and Isometries}}

%{\Huge {\bf The Case of Isometries}}

\vspace{1cm}

{\Large  Mario De Marco,$^{\sharp}$ Shani Nadir Meynet$^{\dagger}$}\\

\vspace{1cm}

{\it
{\small

$^\dagger$ Mathematics Institute, Uppsala University, \\ Box 480, SE-75106 Uppsala, Sweden\\
\vspace{.25cm}

$^\dagger$Centre for Geometry and Physics, Uppsala University \\
Box 480, SE-75106 Uppsala, Sweden\\

\vspace{.25cm}
$^\sharp$ Physique Th\'eorique et Math\'ematique and International Solvay Institutes\\
Universit\'e Libre de Bruxelles, C.P. 231, 1050 Brussels, Belgium
\vspace{.25cm}
\vspace{.25cm}
%$^*$ TODO\\
%\vspace{.25cm}
}}

\vskip .5cm
{\footnotesize \tt mario.de.marco@ulb.be   \hspace{1cm} shani.meynet@math.uu.se}

\vskip 1cm
     	{\bf Abstract }
\vskip .1in

\end{center}

\noindent In this work we consider the relation between finite isometries of the internal space and symmetries of the transverse field theory in Geometric Engineering. On top of the established relation between branes wrapping torsional cycles and topological defects, we study other symmetries of the field theory that are not captured by branes wrapped at infinity. Isometries of the engineering geometry have a representations on the field theory, encoded by their non-trivial actions on exceptional and torsional cycles. In this work we describe such action in general terms. As examples, we focus on three classes of geometric engineering spaces: Du Val singularities, toric Calabi-Yau threefolds and $G_2$ manifolds obtained as fibrations of Du Val singularities over $\mathbb S^3$.  In the latter case, we check explicitly that M-theory on Calabi-Yau or $G_2$ manifold with finite isometries reproduces correctly the higher group structures and non-invertible symmetries of the transverse field theories. 

\eject

\end{titlepage}

% titlepage 
{
  \hypersetup{linkcolor=black}
  \tableofcontents
}

\section{Introduction and Conclusions}

The way in which symmetries are understood in quantum field theory changed dramatically during the last years.\footnote{We redirect the reader to the following recent reviews  \cite{
McGreevy:2022oyu,
Freed:2022iao,
Gomes:2023ahz,
Schafer-Nameki:2023jdn,
Brennan:2023mmt,
Bhardwaj:2023kri,
Shao:2023gho,
Carqueville:2023jhb,
Costa:2024wks} for a extensive discussion on the subject.} Following the original idea of \cite{Gaiotto:2014kfa}, symmetries are implemented via topological operators and defects providing generalised quantum numbers for extended field configurations. One of the most interesting features of generalised symmetries is that their fusion rules go beyond the usual group law. These ideas are tightly related to the concept of \textit{topological symmetry 
theory} (SymTFT), an auxiliary topological field theory that can be used to encode all 
the data of symmetries of a given system of interest, via an interplay of the topological bulk and its 
boundary conditions \cite{Bhardwaj:2017xup,
Chang:2018iay,
Komargodski:2020mxz,
Heidenreich:2021xpr,
Sharpe:2021srf,
Choi:2021kmx,
Kaidi:2021xfk,
Roumpedakis:2022aik,
Bhardwaj:2022yxj,
Gaiotto:2020iye,
Apruzzi:2021nmk,
Freed:2022qnc,
Kaidi:2022cpf,
Kaidi:2023maf,
Baume:2023kkf,
Brennan:2024fgj,
Antinucci:2024zjp,
Bonetti:2024cjk, 
Bhardwaj:2024qiv, 
Hasan:2024aow,
Cordova:2024goh,
Cordova:2024iti,
Decoppet:2024htz
}.

The SymTFT is a powerful tool to encode the data of a field theory and its various global variants. However, a Lagrangian description of the theory of interest is often needed in order to recover it. Unfortunately, it is believed that the vast majority of quantum theories lack a conventional Lagrangian description.

Nevertheless, for models with conserved supercharges, a powerful 
collection of techniques were developed in the context of string/M-/F-theory, where one is able to describe non-Lagrangian theories in terms of geometrical quantities, both in the context of compactifications, holography and geometric engineering. In particular, extensive work has been done in the relation between geometrical data and generalized symmetries \cite{
Tachikawa:2013hya,
DelZotto:2015isa,
Morrison:2020ool,
Albertini:2020mdx,
Closset:2020scj,
Closset:2021lwy,
Apruzzi:2021vcu,
Hosseini:2021ged,
Apruzzi:2021mlh,
Bhardwaj:2021mzl,
GarciaEtxebarria:2022vzq,
Apruzzi:2022rei,
Antinucci:2022vyk,
Hubner:2022kxr,
DelZotto:2022fnw,
DelZotto:2022joo,
Cvetic:2022imb,
Heckman:2022muc,
Heckman:2022xgu,
Bashmakov:2022jtl,
Bashmakov:2022uek,
Bashmakov:2023kwo,
Amariti:2023hev,
Acharya:2023bth,
Dierigl:2023jdp,
Cvetic:2023plv,
Lawrie:2023tdz,
Apruzzi:2023uma,
Yu:2023nyn,
Garding:2023unh, 
Etheredge:2023ler,
Arias-Tamargo:2023duo,
Heckman:2024oot,
Heckman:2024obe,
Argurio:2024kdr,
Huertas:2024mvy,
Cvetic:2024dzu,
Braeger:2024jcj,
Heckman:2024zdo,
Gagliano:2024off,
Najjar:2024vmm,
Cvetic:2025kdn,
Heckman:2025isn
}. The scope of this work is to further explore the relation between symmetries and geometry in the context of 
Geometric Engineering  --- see e.g. \cite{DelZotto:2024tae} for a review. As anticipated, Geometric Engineering provides a 
dictionary between field theoretical quantities and geometrical ones by considering limits of String/M-/F-
theory on a non-compact manifold $X$. In this framework, the excitations of the field theory on the transverse 
directions, $\mathcal T_X$, are given either by branes wrapping exceptional cycles in the geometry, or by 
normalizable mode expansions of the internal metric and of the RR fields. Theories obtained in this way are 
relative, i.e. the geometry does not provide an absolute choice of topological boundary conditions for the 
symmetries background fields due to flux noncommutativity \cite{Freed:2006ya,Freed:2006yc,Moore:2004jv}. 
Luckily, the geometry provides all the necessary data to encode all possible global realizations of various 
symmetries and their anomalies, effectively providing the SymTFT, $\mathcal F_X$, of the dynamical theory. Indeed, the 
SymTFT can be read from the boundary at infinity of the non-compact geometry, $\partial X$, \cite{Apruzzi:2021nmk,GarciaEtxebarria:2024fuk} and its operators can be described as branes wrapping torsional cycles of the boundary \cite{Heckman:2022muc,DelZotto:2024tae}. Different choices of boundary conditions in the 
SymTFT amounts in choosing which set of branes is allowed to end on the boundary at infinity and thus become 
genuine operators of the theory giving the string theory realization of the discussion in \cite{Kaidi:2022cpf}. In this context, the action of 
symmetry operators can be recovered from the linking of membranes. This correspondence was explored deeply in 
a previous work, \cite{DelZotto:2024tae}, where 't Hooft anomalies for higher form symmetries were recovered from higher linking number of membranes. 

This analysis is able to capture discrete symmetries, since it is tightly linked to torsional cycles, but it is not able to see all of them. As we will argue in the rest of the paper, certain 0-form symmetries are realized as discrete isometries of the engineering geometry $X$ and cannot be led back to wrapped branes on $\partial X$.\footnote{See \cref{sec:7dsym} for more details.} In this work, we investigate this relation, studying the induced action of the isometries on the d.o.f. of the field theory, as well as the higher group structure they posses. In addition, we will comment on the realization of these symmetries at the level of the SymTFT. This analysis will also allow us to get hints on the Geometric Engineering description of the non-invertible duality defects of 4D $\mathcal N=1$ SYM from the perspective of M-theory on non-compact $G_2$ manifolds. Moreover, as further consistency checks for our proposal, we exploit the geometric engineering realisation of 7d SYM theories, 4d $\mathcal{N} = 1$ SYM, and 5D SCFTs, commenting, in the latter case, on the relation between the center symmetry and the particle-instanton duality that arises for certain geometries. These results are obtained by leveraging some uniqueness results of the Ricci-flat metric for the manifold of interest, allowing us to study the isometries in terms of the biholomorphisms of the engineering space, which in general are much easier to describe. 

Ultimately, we will argue that the correct topological quantity encoding the data of the SymTFT it is not only the (co-)homology of the boundary geometry, but the (co-)homology with local coefficient, i.e. the boundary (co-)homology twisted by the action of the isometries. This proposal follows from the fact that generic automorphisms of the SymTFT, i.e. 0-form symmetries of the latter, are associated with dualities of the underlying field theory, and when the operators associated to such symmetries admit boundary conditions, they can become genuine 0-form symmetries of the dynamical theory. This observation was already made in the context of field theory and holography \cite{Kaidi:2022cpf, Antinucci:2022vyk, Antinucci:2023ezl}, but, to our knowledge, was never explicitly linked to the internal geometry of the engineering space, and in particular with its isometries. Despite making this observation, we will not delve into the details of how the full SymTFT can be recovered from geometry, as in \cite{Apruzzi:2021nmk}, in terms of twisted cocycles, leaving it as a possible follow-up. On the same line, another interesting direction would be the geometrisation of the different global variants one can realize after gauging symmetries associated to isometries. These operations are less transparent in this context, compared to the field theory one. On general ground, we expect that gauging a discrete symmetry may produce new interesting geometries,\footnote{In  \cite{Acharya:2021jsp, Apruzzi:2022nax} similar ideas where explored, even if in the context of S-fold projection. See also \cite{Arias-Tamargo:2023duo} for comments on the type IIB case.} and possibly leading to the realization of theory with non simply laced flavor groups via geometric engineering, a so far elusive construction in the context of 5d model building.
\bigskip

This paper is organised as follows. In \cref{sec:hilb} we give a brief review of some features of geometric engineering dictionaries we will use in this work. In particular, we revisit the action of finite 0-form symmetries arising from isometries and dualities and their interplay with defect groups. 
In \cref{sec:geometricdefautomorphism} we study the isometries-induced 0-form symmetries in the context of hypersurfaces. We start, in \cref{sec:resduval}, by considering the case of (resolved) Du Val singularities, giving a detailed description of the action of the isometries on exceptional cycles and torsional ones. We then discuss, in \cref{sec:autgroupsurfaces} general properties
of the these 0-form symmetries in the context of geometric engineering on hypersurface singularities.
In \cref{sec:mixedanomalies7dsym} we apply our discussion to 7d SYM (M-theory on Du Val singularities) and 4d $\mathcal N = 1$ SYM (M-theory on specific $G_2$-manifolds). We then comment, in \cref{sec:symtft} on the consequences of isometries symmetries for the SymTFT  and the possibility of realizing non-invertible symmetries as discussed in \cite{Bhardwaj:2022yxj, GarciaEtxebarria:2022vzq}. In \cref{sec:toricCY3} we discuss application to 5d SCFTs obtained by considering M-theory on toric Calabi-Yau threefolds. Finally, in \cref{sec:complexdefs} we comment on the physical consequences of isometries on theories obtained via geometries admitting consistent complex structure deformations.

%\section{Symmetries from geometry: a lightning review}\label{sec:GE}

\section{Symmetries from isometries: a lightning review}\label{sec:hilb}

%In this section we give a lightning review of the geometric origin of 0-form symmetries from isometries in geometric engineering.

%For a more thorough review of geometric engineering dictionaries we refer our readers to the discussion in Section 2 of \cite{DelZotto:2024tae}, whose notation we closely follow.

%In this section, we briefly review the Geometric Engineering (GE) paradigm \cite{Katz:1996fh, Katz:1996th, Bershadsky:1996nh}, with emphasis on how topological operators and higher groups can be read from geometrical data \cite{DelZotto:2015isa,Garcia-Etxebarria:2019caf, Albertini:2020mdx,Morrison:2020ool}. In particular, we describe how finite isometries of the engineering manifold, that act in a non-trivial way on the boundary homology cycles, can be interpreted as 0-form symmetries acting on higher form symmetries leading to higher group structures.

%\subsection{BPS excitations and 0-form symmetries from isometries}

Geometric engineering (GE) methods provide dictionaries between stable BPS backgrounds and supersymmetric field theories of the form
\begin{equation}
\mathcal T_X(M) = GE_{\mathscr S / X}(M) = \mathcal Z_{\mathscr{S}}(M \ltimes X)
\end{equation}
where $M \ltimes X$ is a stable BPS background for string/M/F theory $\mathscr S$, $M$ is a $d$ dimensional compact manifold, $\mathcal T_X(M)$ is the partition function of a $d$-dimensional supersymmetric field theory $\mathcal T_X$ on  $M$, and $X$ is typically a singular BPS geometry (arising from a geometric engineering limit \cite{Katz:1996fh, Katz:1996th, Bershadsky:1996nh} of a string compactification). These dictionaries can be extended to give access to many other features of the $d$-dimensional field theory $\mathcal T_X$. For a review of the aspects relevant for our analysis we refer our readers to \cite{DelZotto:2024tae},  whose notation we closely follow.

\medskip

%In particular, by geometric engineering, the BPS excitations of $\mathcal T_X$ are obtained by wrapping branes on vanishing BPS cycles in $X$. A $p$-dimensional membrane wrapping one such $k$-cycle, give rise to a $p+1-k$ dynamical BPS object (a particle, a string, or a higher dimensional excitation), with an energy density controlled by the volume of the cycle in $X$. 

In this paper we are interested in the case the internal manifold $X$ has an isometry $\Psi$, ie. a map $\Psi: X \to X$ that preserves the metric $\Psi(g) = g$.\footnote{In particular, when $X$ is K\"ahler, a biholomorphism is an isometry if it leaves the K\"ahler form invariant, since $\Psi(g_{i \bar{j}}) = \Psi(\omega_{i \bar{k}})I^{\bar k}_{\bar j}$, with $\Psi$ acting trivially on $I$, the complex structure.} In general, the set of all $\Psi$ form a (Lie) group, $\mathscr I(X,g)$. From geometric engineering, we expect that $\mathscr I(X,g)$ induces non-trivial transformations on the BPS excitations. This occurs because in geometric engineering, the charge lattices are encoded in homology $H_k(X,\mathbb Z)$ --- a $(p+1-k)$-dimensional BPS excitation arising from a $p$-brane wrapping a vanishing $k$-cycle $S$ has charge $[S] \in H_k(X,\mathbb Z)$ with respect to the $U(1)$ higher form gauge field arising from the harmonic decomposition of the corresponding higher $(p+1)$-form potential. $\mathscr I(X,g)$ acts on the charge lattice via pushforwards
\begin{equation}
    \Psi_*: [S] \in H_\bullet(X,\mathbb Z) \to [\Psi_{*}(S)] \in H_\bullet(X,\mathbb Z) \, ,
\end{equation}
which therefore gives representations of $\mathscr I(X,g)$ on the whole spectrum of the theory. As discussed in \cite{DelZotto:2024tae}, it is natural to interpret $\Psi \in \mathscr I(X,g)$ in terms of a 0-form symmetry topological operator. Such operators can act on field configurations of any dimensionality by crossing them, and on point-like operators by linking them. 

\begin{figure}
\begin{center}
\begin{tabular}{p{5cm}cp{7cm}} 
$\begin{gathered}
\begin{tikzpicture}
    \draw [->] (0,0) -- (0,3);
        \node[left] at (0,3){$ \mathbb R_\text{time}$};
        \draw[fill,blue] (0,1.5) circle (0.05cm);
        \node[left] at (-0.3,1.5){$t_0$};
        \node[blue] at (3.1,1.5){$U_\Psi$};
        \draw[fill,color=blue!50] (1.6,1.5) ellipse (1cm and 0.3cm);
    \draw (1.6,0.5) ellipse (1cm and 0.3cm);
    \node at (1.6,0.5){$N$};
    \draw[dashed] (0.6,0) -- (0.6,3);
    \draw[dashed] (2.6,3) -- (2.6,0);
\end{tikzpicture}
\end{gathered}
$
&$\longleftrightarrow$&
$\begin{gathered}\begin{tikzpicture}
    \draw [->] (0,0) -- (0,3);
        \node[left] at (0,3){$ \mathbb R_\text{time}$};
                \draw[fill,blue] (0,1.5) circle (0.05cm);
        \node[left] at (-0.3,1.5){$t_0$};
         \draw (1.6,0.5) ellipse (1cm and 0.3cm);
    \node at (1.6,0.5){$N$};
    \draw[dashed] (0.6,0) -- (0.6,3);
    \draw[dashed] (2.6,3) -- (2.6,0);
    \draw (3,0.5) -- (5,0.7);
    \draw (3,0.5) -- (5,0.3);
    \node[right] at (5,0.5){$X$};
    \draw[fill,color=blue!50] (3,1.5) -- (5,1.3) -- (5,3) -- (3,3);
    \node[right] at (5,1.5){$\Psi(X)$};
    \draw (3,1.5) -- (5,1.7);
    \draw (3,1.5) -- (5,1.3);
    \draw[dashed] (3,3) -- (3,0);
\end{tikzpicture}
\end{gathered}
$\\
\end{tabular}
\caption{On the left: The topological operator $U_{\Psi}$ inserted at time $t_0$, filling the spatial directions, corresponds to a unitary operators acting on the Hilbert space associated to the spatial slice $N$. On the right: The action of the isometry on the geometric background. These two actions are related by the geometric engineering dictionary.}\label{fig:ISOaction}
\end{center}
\end{figure}
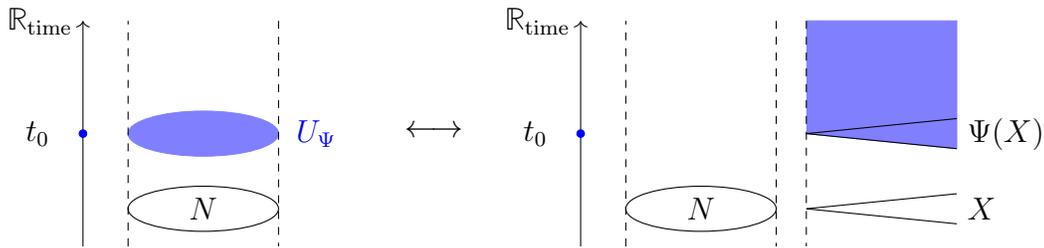

To geometrically engineer the action of $\Psi$, we consider a geometric background of the form $(\mathbb R \times N) \ltimes X$. In this context, $\mathbb R$ is the time direction and at each point along this line we can associate an Hilbert space, $\mathcal T_X(N)$, via Hamiltonian quantization. The action of $\mathscr I(X,g)$ on $\mathcal T_X(N)$ arises from backgrounds of the form in Figure \ref{fig:ISOaction}, where at $t_0$, $(X,g)$ and $(\Psi(X),\Psi(g))$ are identified via the action of the isometry.

Let us note that the presence of isometries depends explicitly on the different choices of moduli of the engineering geometry, let us denote such space $\mathcal M$. Indeed, it may happen that an isometry is present only for specific choices of moduli $m_* \in \mathcal M$. As an example, consider a geometry which has a torus fibration. When the torus parameter is tuned to $\tau = i$ one has a $\mathbb Z_4$ isometry generated by the modular transformation $S$ acting by swapping the A and B cycles, see \cite{DelZotto:2024tae} for an example.  
The space $\mathcal M$ is interpreted as extended moduli spaces for the theories $\mathcal T_X$, where points in $\mathcal M$ corresponds to choices of scalar VEVs and parameters (corresponding, respectively, to normalizable and non-normalizable moduli in the geometric engineering limit). The fact that the isometry is broken for generic choices of moduli resonates with the fact that in field theory a 0-form symmetry can be (spontaneously) broken by generic choices of scalar VEVs.

To make the above statement more clear, let us consider a vacuum of the field theory, labeled by a specific choice of moduli, $m \in \mathcal M$, by $\ket{m}$. The action of the isometry on the geometry induces an action on $\mathcal M$, $\Psi: \mathcal M \to \mathcal M$. This operator acts on the vacuum as
\begin{equation}
    U_{\Psi} \ket{m} = \ket{\Psi(m)} \, .
\end{equation}
We then have that
\begin{align}\label{eq:topopinsertion}
    \braket{\Psi(m)}{m} = \bra{m} U_{\Psi} \ket{m}
\end{align}
is either 0 or 1, up to normalizations, according to whether the symmetry is broken or not. In particular, if $m = \Psi(m)$, i.e. the moduli are the same before the action of the biholomorphism of the internal geometry, $\Psi$ is an isometry and the operator $U_{\Psi}$ is a symmetry of the vacuum. 

The geometric counterpart of \cref{eq:topopinsertion}, depicted in \cref{fig:ISOaction}, corresponds to the insertion of the operator $U_{\Psi}$ along the  spatial slice $t=t_0$ in the field theory. From the point of view of the internal geometry, this corresponds to consider the space to be $X_m$ for positive values of $t$ and $\Psi(X_m)$ for negative ones.\footnote{By $X_m$ we mean the crepant resolution of $X$ corresponsing to the field theory vacuum $\ket{m}$.}

If we have $\Psi(X_m) \neq X_m$ the corresponding 
internal geometry should display some sort of discontinuity. We expect that this phenomenon presents itself via divergencies of the Riemann tensor at $t = 0$ in the form of derivatives of a Dirac delta.\footnote{This can be argued by noticing that the Riemann tensor is 
proportional to the second derivative of the metric and if $
\Psi$ is not an isometry, then the metric on the internal space 
is proportional to an Heaviside $\Theta$ (as it depends linearly 
on the $m$, that are sent to $m' \neq m$ by $
\Psi$). We thank Alessandro Tomasiello for explaining this point 
to us.} Such a solution looks extremely hard to realize from the 
supergravity perspective (and thus for the geometric engineering counterpart as well), thus suggesting that the operator $U_{\Psi}$ becomes a domain wall between two vacua. On the other hand, if $X_m = \Psi(X_m)$, the corresponding Riemann tensor profile is smooth at $t = 0$, and hence the solution is allowed even from a geometric perspective, therefore we expect that \cref{eq:topopinsertion} is non-vanishing also from a gravitational point of view.

In the context of this work, we will mainly considered geometries built using Du Val surface singularities $X_{\mathfrak g}$. We have different cases:  
\begin{enumerate}
    \item In the case of 7d SYM theories studied in \cref{sec:7dsym}, as well as in the 5d SCFTs of \cref{sec:toricCY3}, the isometry is present only when, in a phase associated with a resolved or deformed singularity, the cycles exchanged by $\Psi$, have the same volume.  
  
    \item In the case of 4d $\mathcal N = 1$ SYM, that we study in \cref{sec:4dnequal1}, the isometry is always present. Indeed, the $G_2$ manifold used for engineering has no tunable parameter, but the volume of the base $S^3$, which does not break the isometry of the fibered Du Val.

    \item In the case of 6d $(2,0)$ theories, that we study in \cref{sec:6d20full}, we have (similarly to the 7d case) that the presence of the isometry will depend on the choice of complex structure deformation of the Du Val singularity used to engineer the theory.
\end{enumerate}

Before moving on, let us stress two points: first, more generally, a diffeomorphism of the engineering manifold may not preserve all moduli, i.e. $X_m \neq X_m'$, but the two physical theories may be related by a duality. This fact was commented in \cite{DelZotto:2024tae}, where the $SL(2,\mathbb{Z})$ duality of pure $\mathcal{N}=4$ SYM could be engineered in type IIB on a Du Val singularity times a torus. In the case of pure SYM, only when the torus has a specific shape the generic diffeomorphism becomes an isometry, however, such a transformation may still be a duality. This is reminiscent of the Class $\mathcal{S}$ construction, in which the mapping class group of the Riemann surface is interpreted as the duality group of the associated field theory, and dualities become symmetries only when an element of the mapping class group is an automorphism of the surface.

Second, we do not claim that all 0-form symmetries of the theory $\mathcal T_X$ can be captured by 
isometries of $X$. Indeed, continuous 0-form symmetries can be 
identified from bulk-boundary systems arising from higher 
codimension singular loci of non-isolated singularities, 
\cite{Hayashi:2019fsa,DelZotto:2022joo,Acharya:2023bth,Heckman:2024oot,Bonetti:2024cjk}. However, the role of 
isometries is often neglected in the context of GE, thus our 
goal is to clarify the interplay between isometries and 
symmetries of field theories. In the following we will focus on 
some specific examples of theory $\mathcal T_X$, where $X$ is 
taken to be a Calabi-Yau two/threefold or a $G_2$ singular cone, 
and in which we have isometries corresponding to finite 0-form 
symmetries of $\mathcal T_X$.

\subsection{Heavy defects and defect groups}\label{sec:defhilb}

In order to understand how the isometries may induce an action on the higher form symmetries of the theory, we first review how “heavy" defects, such as Wilson/'t Hooft lines/surfaces, are realized in GE and their relation to the homology of $X$.

Heavy defects of dimension $(p-k+1)$ are obtained by choosing a non-compact $k$-dimensional cycle in relative homology
\begin{gather}
    S \in H_k(X,\partial X)
\end{gather}
and considering a $p$-brane wrapped on it. Since the cycle is non-compact, the resulting BPS state has infinite mass. We denote such probe defect by
\begin{gather}\label{eq:wilson} 
     \mathcal W_\textbf{Dp}^S \, ,
\end{gather}
these heavy defects can be thought as Wilson / 't Hooft operators of various dimensions.

Now, dynamical BPS states can break the heavy defects just described in a generalised version of the 't Hooft screening mechanism. Geometrically, this condition can be written as
\begin{equation}
[S] \in H_k(X,\partial X) / H_k(X) \, ,
\end{equation}
where $[S] \neq 0$ is a necessary condition for the stability of the heavy defect. If $[S] = 0$, the defect is \textit{endable}, meaning that (at sufficiently high energies) there are BPS excitations that can cause its screening. This motivates the definition of the \textit{defect group} for $\mathcal T_X$ as  \cite{DelZotto:2015isa,Garcia-Etxebarria:2019caf,Albertini:2020mdx,Morrison:2020ool,Hubner:2022kxr}
\begin{equation}
\begin{aligned}\label{eq:defectgroup}
        \mathbb{D}_X := \bigoplus_n \mathbb{D}^{(n)}_X \quad \text{where}\ \  \mathbb{D}^{(n)}_X =  \bigoplus_{p\text{-branes}}\left(\bigoplus_{k \text{ s.t. } \newline p-k+1=n} \left(\frac{H_k(X , \partial X)}{H_{k}(X)}\right)\right)\, .
\end{aligned}\end{equation}
Elements of $\mathbb{D}^{(n)}_X$ correspond to charges of $n$-dimensional defects that cannot be screened.

Since these heavy defects carry a charge under higher form symmetries, it is natural to ask whether or not one can engineer topological defects that do measure these charges. The answer is positive as we discuss now briefly.

To simplify the discussion of our examples, we assume that the theory $\mathcal T_X$ is a superconformal field theory (SCFT). Then, the engineering $l+1$-dimensional manifold is a conical singularity with special holonomy
\begin{equation}
\label{eq:conedefinition}
X = \mathrm{Cone}(\mathbf{L}_X),\, \qquad \mathrm{d} s^2_X = \mathrm{d} r^2 + r^2 \mathrm{d} s^2_{\mathbf{L}_X}.
\end{equation}
The space $ L_X$ is the \textit{link} of the singularity.\footnote{\ It is a $l$-dimensional compact manifold obtained by intersecting the given singularity with a sphere centered on it. If $X$ has special holonomy, the corresponding link has a metric with special properties inherited from the special holonomy of $X$. For example, if $X$ is a Calabi-Yau singularity with special holonomy $SU(n)$, the resulting link is a $(2n-1)$-dimensional manifold with a Sasaki-Einstein metric (possibly with orbifold singularities).} 

We can construct extended topological operators by wrapping branes on the torsional cycles in the link
\begin{equation} 
    \beta \in H_k(\mathbf{L}_X)
\end{equation} and on compact cycles $\Sigma$ in the $d$ dimensional spacetime as
\begin{equation}\label{eq:topodefecto} 
    \mathcal D_{\mathbf{Mp}}^\beta(\Sigma) 
\end{equation} to distinguish them from the ones giving rise to defects $\mathcal W_{\mathbf{Mp}}^S$, that we have introduced in \cref{eq:wilson} in the previous section. 

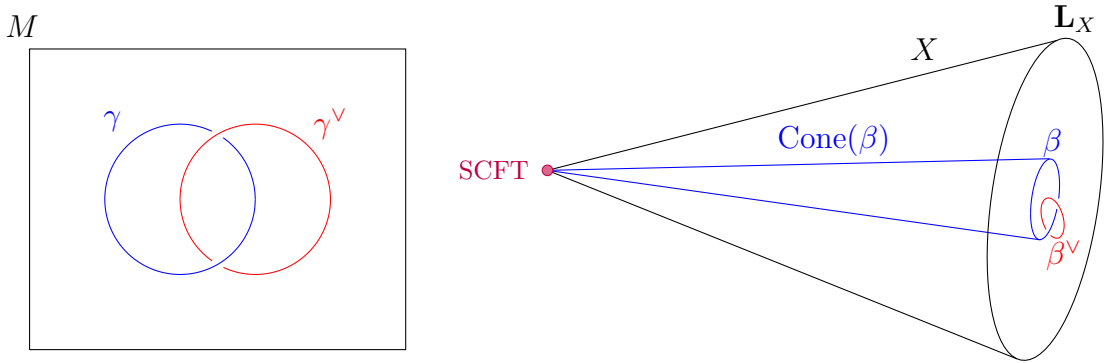
\begin{figure}
    \centering
    \begin{tikzpicture}
        \draw[black] (-4,0,-1) -- (3.05,2,-0.3);
        \draw[black] (-4,0,-1) -- (2.75,-2,0.32);
        \draw[blue] (3,0,0.5)
            \foreach \t in {5,10,...,145}
                {--(3,{-0.5*sin(\t)},{0.5*cos(\t})}
            ;
        \draw[blue] (3, -0.17101, -0.469846)
        \foreach \t in {165,...,355}
                {--(3,{-0.5*sin(\t)},{0.5*cos(\t})}
        -- (3,0,0.5);
        
        \draw[blue] (-4,0,-1) -- (3.06,0.54,0);
        \draw[blue] (-4,0,-1) -- (2.94,-0.54,0);
        
        \draw[black] (3,0,2)
            \foreach \t in {5,10,...,355}
                {--(3,{-2*sin(\t)},{2*cos(\t})}
            -- (3,0,2);
        
        \draw[red] (3.1,0,0) 
            \foreach \t in {5,10,...,190}
                {--({3.1+0.25*sin(\t)},{-0.25+0.25*cos(\t)},{0.25*sin(\t)})}
                ;
        \draw[red] (2.95661, -0.454788, -0.143394)
            \foreach \t in {220,...,355}
                {--({3.1+0.25*sin(\t)},{-0.25+0.25*cos(\t)},{0.25*sin(\t)})}
            -- (3.1,0,0) ;
        \draw[purple, fill=purple!60] (-4,0,-1) circle (2pt);
        \node[left, purple] at (-4.1,0,-1){{\footnotesize SCFT}};
        \node[blue] at (0,0.6,-0.5){$\mathrm{Cone}(\beta)$};
        \node[blue] at (3.1,0.75,0){$\beta$};
        \node[red] at (3.25,-0.75,0){$\beta^\vee$};
        \node[black] at (3.4,2.4,0){$\mathbf{L}_X$};
        \node[black] at (1.4,2.0,0){$X$};

        \draw[black] (-10.5,-2,0) -- (-5.5,-2,0) -- (-5.5,2,0) -- (-10.5,2,0) -- (-10.5,-2,0);
        
        \draw[blue] (-7.5,0,0) 
            \foreach \t in {5,10,...,55}
                {--({-8.5+cos(\t)},{sin(\t)},0)}
            ;
        \draw[blue] (-8.07738, 0.906308)
            \foreach \t in {70,...,355}
                {--({-8.5+cos(\t)},{sin(\t)},0)}
        --(-7.5,0,0);
        \draw[red] (-8.5,0,0)
            \foreach \t in {5,10,...,55}
                {--({-7.5-cos(\t)},{-sin(\t)},0)}
            ;
            
        \draw[red] (-7.92262, -0.906308)
            \foreach \t in {70,...,355}
                {--({-7.5-cos(\t)},{-sin(\t)},0)}
        --(-8.5,0,0);
        \node[black] at (-10.6,2.3,0){$M$};
        \node[blue] at (-9.4,1.04,0){$\gamma$};
        \node[red] at (-6.5,1,0){$\gamma^\vee$};
    \end{tikzpicture}
    \caption{Schematic description of geometric engineering on a conical singular geometry. }
    \label{fig:cone}
\end{figure}

In particular, the charges of $\mathcal T_X$ are organised via the long exact sequence in relative homology
\begin{equation}
\cdots \to H_{k+1}(X) \to H_{k+1}(X,\mathbf{L}_X) \to H_{k}(\mathbf{L}_X) \to H_{k}(X) \to \cdots.
\end{equation}
In all the examples we consider in this paper, this long exact sequence truncates, which allows to easily read off the charges of extended operators. Consider for example the case $H_{k}(X) = 0$ above. For a defect $\mathcal W_{\mathbf{Mp}}^S$ of $\mathcal T_X$ with $S = \text{Cone}(\beta) \in H_{k+1}(X,\mathbf{L}_X) $ such that $[S] \neq 0$ in the defect group, there is a topological membrane, namely $\mathcal D^{\beta^\vee}_{\mathbf{Mp}}$, obtained by wrapping a $k$-cycle $\beta^\vee \in H_k(\mathbf{L}_X)$ such that the two cycles in $\mathbf{L}_X$ do link. This configuration is illustrated in blue in \cref{fig:cone}.

\medskip
  
\subsection{Bulk-boundary cycles relations}
\label{sec:bulk-boundarycycles}
The last piece of the puzzle we need to solve is how to relate the action of the isometry on the exceptional cycles, responsible for the BPS states, to the boundary cycles, responsible for the higher form symmetry operators. This relation can be easily described using the exact sequence in relative homology (that we reproduce for convenience):
\begin{equation}
\cdots \to H_{k+1}(X) \xrightarrow{g} H_{k+1}(X,\mathbf{L}_X) \xrightarrow{f} H_{k}(\mathbf{L}_X) \to 0 \, ,
\end{equation}
where we assume $H_k(X)=0$ for $k$ odd (as it is the case for Calabi-Yau 2/3-folds that we consider in this paper).

Now, given a torsional cycle $ \alpha_k \in H_{k}(\mathbf{L}_X)$, satisfying $n \alpha_k = 0 = \partial \beta_{k+1}$, we can associate a cycle $\kappa_{k+1} \in H_{k+1}(X,\mathbf{L}_X)$ using the fact that $f$ is surjective, i.e. $f(\kappa_{k+1}) = \partial \beta_{k+1}/n = \alpha_k$, we can now pick $z_{k+1} \in  H_{k+1}(X)$ such that $g(z_{k+1}) = n \kappa_{k+1}$, which is in ker$f$. Now it is easy to see that we have the following $\alpha_{k} = f ( g (  z_{k+1} )/n) $.\footnote{See \cite{Apruzzi:2021nmk} for a discussion on this correspondence between torisonal cycles and compact ones.}

This fact will become crucial in the following. Indeed, as we will see, it is possible to express the torsional cycle associated to generator of the higher form symmetry in terms of a linear combination of compact cycles. Knowing how isometries act on compact cycles will allow us to deduce the action of the 0-form symmetry on the higher ones. This action can be described in term of 2-groups or higher generalizations \cite{Benini:2018reh}. We will provide explicit examples of this fact in the case of Du Val singularities, toric CY$_3$ and certain class of G$_2$ manifolds.

\section{Singular hypersurfaces, biholomorphisms and isometries}
\label{sec:geometricdefautomorphism}

In this section we analyze the isometries of hypersurface singularities by studying their biholomorphisms. The relation between these geometrical transformations is obtained, in the Du Val case, by advocating some rigidity results on the associated K\"ahler metric. We begin, in \cref{sec:resduval}, by studying the discrete isometries of the Du Val singularities $Y_{\mathfrak g}$, reviewing the resolution of $Y_{\mathfrak g}$ and making explicit the action on the exceptional and torsional cycles of the biholomorphisms/isometries that we want to relate to 0-form symmetries. Then, we comment on the general properties of isometries and biholomorphism for hypersurface singularities in \cref{sec:autgroupsurfaces}, highlighting the relevant features for the study of 0-form symmetries of the associated theories. We recall that biholomorphisms are holomorphic maps of the GE space to itself, that admit an holomorphic inverse. We will use instead the word "automorphisms" to refer to the outer-automorphisms of the Lie algebra associated with the Du Val singularity.

\subsection{Resolution of Du Val singularities and biholomorphisms}
\label{sec:resduval}
In order to keep track of the action of the isometries on the resolution of $A_{j}, D_{2j+1}, D_{2j+2}, E_6$ singularities, we present now a review of how to resolve these singularities using the the {\it toric ambient space} approach.

The biholomorphisms of interest of the Du Val surfaces are \cite{Vafa:1997mh}: 
\begin{align}
\label{eq:vafaaut}
A_{2j-1}& &x y - z^{2j} = 0 \qquad \qquad \left\{x \leftrightarrow y, z \to - z\right\} \nonumber \\
A_{2j}& &x y - z^{2j+1} = 0 \qquad \qquad \left\{x \leftrightarrow y, z \to z\right\} \nonumber \\
D_{j}& &x^2 + z y^2 + z^{j-1} = 0 \qquad \left\{x \leftrightarrow -x, y \to - y\right\} \nonumber \\
E_{6}& &x^2 +  y^3 + z^{4} = 0 \qquad \quad \left\{x \leftrightarrow -x, z \to - z\right\} \nonumber \\
\end{align}
The equations in the left column of \cref{eq:vafaaut} define the Du Val singularities isolated hypersurfaces of $\mathbb C^3$. The resolution inflates a set of $\mathbb P^1$s intersecting according to the Dynkin diagrams of the ADE algebra associated to each singularity, as depicted in \cref{tab:DynkinADE}. As we will show shortly, biholomorphisms of the Du Val polynomial can be associated to the automorphisms of the corresponding Dynkin diagram. 
\begin{table}
\centering 
    \begin{tabular}{c|c}
        $Y_{A_j}$ &  $\overbrace{\begin{matrix}
\scalebox{1.5}{
\begin{tikzpicture}
    
    \draw (0.1,1.5)--(0.9,1.5);
    \draw (1.1,1.5)--(1.9,1.5);

    \draw (0,1.5) circle (0.1); 
    
	\draw (1,1.5) circle (0.1); 
    
	\draw (2,1.5) circle (0.1); 
	\draw (3,1.5) circle (0.1);
  \node a t (0,1.5) {
 \scalebox{0.4}{$1$}};
   \node a t (1,1.5) {
 \scalebox{0.4}{$2$}};
   \node a t (2,1.5) {
 \scalebox{0.4}{$3$}};
    \node a t (3,1.5) {
 \scalebox{0.4}{$n$}};
   \node a t (2.5,1.5) {$\cdots $};
 \node a t (2.5,1.5) {$\cdots $};
\end{tikzpicture}
}
        \end{matrix}}^{\mbox{\tiny $j$ nodes}}$ \\[2px]
        $Y_{D_k}$ &   $\underbrace{\begin{matrix}
 
 \scalebox{1.4}{           \begin{tikzpicture}
    
    \draw (0.1,1.5)--(0.9,1.5);
    \draw (1.1,1.5)--(1.9,1.5);
    \draw (3.1,1.5)--(3.9,1.5);
    \draw (1,1.6)--(1,2.4);

    \draw (0,1.5) circle (0.1); 
    
	\draw (1,1.5) circle (0.1); 
    
	\draw (2,1.5) circle (0.1); 
	\draw (4,1.5) circle (0.1);
	\draw (1,2.5) circle (0.1);
   \node a t (0,1.5) {
 \scalebox{0.4}{$1$}};
   \node a t (1,2.5) {
 \scalebox{0.4}{$2$}};
   \node a t (1,1.5) {
 \scalebox{0.4}{$3$}};
   \node a t (2,1.5) {
 \scalebox{0.4}{$4$}};
    \node a t (4,1.5) {
 \scalebox{0.4}{$k$}};
 \node a t (2.5,1.5) {$\cdots $};
\end{tikzpicture}  
}
        \end{matrix}}_{\mbox{\tiny $j-1$ nodes}}$ \\[2px]
        $Y_{E_6}$ &  $\begin{matrix}
\scalebox{1.4}{            \begin{tikzpicture}
    
    \draw (0.1,1.5)--(0.9,1.5);
    \draw (1.1,1.5)--(1.9,1.5);
    \draw (2.1,1.5)--(2.9,1.5);
    \draw (3.1,1.5)--(3.9,1.5);
    \draw (2,1.6)--(2,2.4);

    \draw (0,1.5) circle (0.1); 
    
	\draw (1,1.5) circle (0.1); 
    
	\draw (2,1.5) circle (0.1); 
	\draw (3,1.5) circle (0.1); 
	\draw (4,1.5) circle (0.1);
	\draw (2,2.5) circle (0.1); 
    \node a t (0,1.5) {
 \scalebox{0.4}{$1$}};
    \node a t (1,1.5) {
 \scalebox{0.4}{$2$}};
    \node a t (2,1.5) {
 \scalebox{0.4}{$3$}};
     \node a t (3,1.5) {
 \scalebox{0.4}{$4$}};
     \node a t (4,1.5) {
 \scalebox{0.4}{$5$}};
     \node a t (2,2.5) {
 \scalebox{0.4}{$6$}};
 
\end{tikzpicture}
}
        \end{matrix}$\\
    \end{tabular}  
    \caption{Intersection pattern among the exceptional $\mathbb P^1$ of the resolution of Du Val singularities whose Dynkin diagram posses non-trivial outer-automorphisms. }
    \label{tab:DynkinADE}
\end{table}

Before moving to the examples, let us notice that all the singularities appearing in \cref{eq:vafaaut} enjoy as well a continuous group of biholomorphism, acting via a quasi-homogeneous rescaling of the ambient space coordinates: 
\begin{equation}
    \label{eq:quasihomgeneral}
    x \to \beta^{w_x}x, \quad y \to \beta^{w_y}y, \quad  z \to \beta^{w_z}z, 
\end{equation}
for $\beta \in \mathbb C^*$ a constant, and the quasi-homogeneous weights $w_{x}, w_{y}, w_{z} \in \mathbb Q$ chosen in such a way to preserve the Du Val singularities, \cref{eq:vafaaut}. However, this transformation does not act on the exceptional cycles of the resolved singularity, as we will prove in full generality in \cref{sec:autgroupsurfaces}.

\subsubsection{\texorpdfstring{$A_j$}{} singularities}
We start by considering the $A_j$ singularity, 
\begin{equation}
    \label{eq:ajequation}
    xy  = z^{j+1}. 
\end{equation}
The singularity is toric, as it is identified by the vanishing locus of a binomial, and admits a toric crepant resolution. The toric fan of the resolved surface is

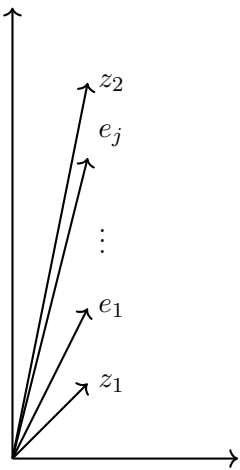
\begin{figure}[H]
\centering
    \scalebox{1.}{
  \begin{tikzpicture}
        \draw[thick,->]  (0,0)--(0,1)--(0,2)--(0,3)--(0,4)--(0,5)--(0,6);
        \draw[thick,->]  (0,0)--(1,0)--(2,0)--(3,0);
        \draw[thick,->]  (0,0)--(1,1);
        \draw[thick,->]  (0,0)--(1,2);
        \draw[thick,->]  (0,0)--(1,4);
        \draw[thick,->]  (0,0)--(1,5);
        \node[right] at (1,1) {\small $z_1$};
        x\node[right] at (1,2) {\small $e_1$};
        \node[right] at (1,3) {\small $\vdots$};
        \node[right] at (1,4.3) {\small $e_{j}$};
        \node[right] at (1,5) {\small $z_2$};
    \end{tikzpicture}}
     \caption{Toric diagram of the (resolved) $A_j$ singularity.}
    \label{fig:Z2sym}
\end{figure}
where the singular surface is obtained by erasing all the internal rays. We will call $e_{k}$, with $k = 1,..., j$ the homogeneous toric coordinates associated to the compact toric divisors $e_{k} = 0$. This set of exceptional $\mathbb P^1$ is intersecting according to the $A_j$ Dynkin diagram in \cref{tab:DynkinADE}. We call $z_{1}, z_2$ the homogeneous coordinates associated to two non-compact divisors, topologically isomorphic to two half-cigars, intersecting the outermost nodes of the Dynkin diagram in two points.

We can now express the singularity in terms of the original coordinates $(x,y,z)$ via the blowdown map
\begin{equation}
    \label{eq:blowudownaj}
    x = e_{1}^1... e_{j}^{j} z_{2}^{j+1}, \quad  y = z_{1}^{j+1}e_{1}^j... e_{j}^{1}, \quad z = z_1 e_1 e_2...e_j z_2.
\end{equation}
The isometry acts by exchanging $x$ and $y$ and sending $z$ to minus itself. This action corresponds to exchanging the exceptional cycles as dictated by the outer automorphism of the associated Dynkin diagram. We thus have, for $A_{2l+1}$,
\begin{equation}
    \label{eq:outerauta2jminusone}
    z_{1} \leftrightarrow z_{2}, \quad e_{k} \leftrightarrow e_{2l+2-k}, \quad e_{l+1} \to - e_{l+1}
\end{equation}
with $k \neq l+1$. For the $A_{2l}$ we have 
\begin{equation}
    \label{eq:outerauta2j}
    z_{1} \leftrightarrow z_{2}, \quad e_{k} \leftrightarrow e_{2l+1-k}
\end{equation}

In particular, calling $\mathbb P^1_k$ the curve identified by $e_k = 0$, we have that the induced action on the $H_{2,cpct}(Y_{A_j},\mathbb Z)$ is
\begin{equation}
\label{eq:inducedactionhomology}
[\mathbb P^1_{k}] \leftrightarrow [\mathbb P^1_{j-k+1}], 
\end{equation}
with $[...]$ the homology class represented by the $\mathbb P^1$ and, for $j = 2l -1$,
\begin{equation}
    [\mathbb P^1_{\ell}] \to [\mathbb P^1_{\ell}]
\end{equation}
At this stage, one might be concerned that this is too fast, and the biholomorphism of the resolved Du Val singularities changes orientation to the permuted $\mathbb P^1$s (giving subtle $\pm$ signs in the pushforward action on the $H_{2}(Y_{\mathfrak g},\mathbb Z)$). However, for a  biholomorphism $\Psi$ we have
\begin{enumerate}
    \item if $\Psi$ exchanges $\mathbb P_i, \mathbb P_j$, then $\Psi \rvert_{\mathbb P_i} \in \text{Hom}(\mathbb P_i,\mathbb P_j)$ is an isomorphism of complex manifolds, and complex isomorphism preserve the orientation\footnote{For example, one can easily convince themselves that any complex map $z \to f(z) = z'$ preserve the sign of the usual Fubini-Study K\"ahler volume form $\omega = \frac{d z \wedge d \overline{z}}{1 + (z \overline{z})^2}$};
    \item if $\mathbb P_i$ is sent to itself by $\Psi$ then $\Psi \rvert_{\mathbb P_i}$ is a biholomorphism of $\mathbb P_i$, and hence preserves its orientation\footnote{This is true despite of the minus sign in the last equation of \cref{eq:outerauta2jminusone}.}.
\end{enumerate}
We will use this same argument also for the other type of Du Val singularities, to rule out non-trivial signs in the pushforward action $\Psi_{\ast}: H_{2}(Y_{\mathfrak g},\mathbb Z) \to H_{2}(Y_{\mathfrak g},\mathbb Z) $.
Going back to the $A_j$ singularity, we notice that, at the level of \cref{eq:ajequation}, the biholomorphism is equivalent to the exchange \begin{equation}
    \label{eq:automorphismsingularaj}
    x \leftrightarrow y, \quad z \to -z.  
\end{equation}
We notice that we can diagonalize the action writing $a = x + y, b = x-y$, getting $b \to -b$ and $a \to a$.

Finally, we note that this biholomorphism is an isometry only if the volumes of the exceptional cycles exchanged by it are the same. We will comment later on the physical consequences of this observation, but for now we just point out that only for specific resolutions a biholomorphism of the singularity is an isometry.

\subsubsection{\texorpdfstring{$D_{4}$}{} singularity}
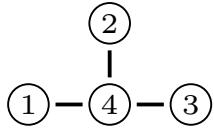
\begin{figure}[H]
\centering
    \scalebox{1.8}{\begin{tikzpicture}
        \draw (-2.8,3.8) circle (0.15);
        \draw[thick] (-2.6,3.8)--(-2.4,3.8);
        \draw (-2.2,3.8) circle (0.15);
        \draw[thick] (-2,3.8)--(-1.8,3.8);
        \draw (-1.6,3.8) circle (0.15);

        \draw[thick] (-2.2,4.0)--(-2.2,4.2);
        \draw (-2.2,4.4) circle (0.15);
        \node at (-2.8,3.8) {\tiny$1$};
        \node at (-2.2,3.8) {\tiny$4$};
        \node at (-1.6,3.8) {\tiny$3$};
        \node at (-2.2,4.4){\tiny$2$};
        \end{tikzpicture}
        }
        \caption{$D_4$ Dynkin diagram; the outer automorphisms act permuting the nodes 1,3,4.}
                \label{fig:dynkinD4}
        \end{figure}

Let us now consider the $D_{j+1}$-series. The case $j = 3$ is peculiar, because for this instance the $\mathbb Z_2$ automorphism group of the Dynkin diagram (shown in \cref{fig:dynkinD4})  is enhanced to $S_3$.\\ \indent 
Let's start from $D_4$, the singularity is 
\begin{equation}
\label{eq:d4singular}
    x^2 - s t (s+t) = 0.
\end{equation}
The resolution inflates a pattern of $\mathbb P^1$s intersecting according to the $D_4$ Dynkin diagram and can be described as \cite{Collinucci:2020jqd} the following hypersurface in the toric ambient space 
\begin{equation}
    \label{eq:d4resolved}
    \sigma w^2 = \lambda s_1 t_1 (s_1+t_1) \qquad \subset \qquad 
\renewcommand{\arraystretch}{1.3}
\begin{array}{ccccc}
\hline
 s_1 & t_1 & w & \lambda & \sigma \\
\hline
 1 & 1 & 1 & -1 & 0   \\
 0 & 0 & 1 & 1 & -1   \\
\end{array}
\end{equation}
with the Stanley-Reissner (SR) condition being 
\begin{equation}
    \label{eq:srideald4}
    (s_1,t_1,w) \neq 0 \qquad \text{and} \qquad  (w, \lambda) \neq 0. 
\end{equation}
The blowdown map, in homogeneous coordinates, reads 
\begin{equation}
    \label{eq:blowdownd4}
    x = w \sigma^2 \lambda, \quad y = s_1 \sigma \lambda, \quad z = t_1 \sigma \lambda
\end{equation}
and four exceptional $\mathbb P^1$s are described as 
\begin{eqnarray}
\label{eq:curvesd4}
&&\mathcal C_{1} \equiv \left\{\sigma = s_1 =
 0\right\} \nonumber \\
&& \mathcal C_{2} \equiv \left\{\sigma = t_1 =  
 0\right\} \nonumber \\
&& \mathcal C_{3} \equiv \left\{\sigma = s_1 + t_1 =
 0\right\} \nonumber \\
&& \mathcal C_{4} \equiv \left\{\sigma = \lambda =
 0\right\} \nonumber \\
\end{eqnarray}
By using \cref{eq:srideald4} and \cref{eq:d4resolved}, we can see that $\mathcal C_{4}$ intersects all the other nodes in a point, and $\mathcal C_{k} \cap \mathcal C_{l} = 0$ if $k \neq 4$ and $l \neq 4$.
The triality group is generated by the three biholomorphisms that (respectively) exchange $s_1 \leftrightarrow t_1$, $s_1 \leftrightarrow t_1 + s_1$, $t_1 \leftrightarrow t_1+s_1$. This action translates into the same permutations of the $(s,t)$ variables after the blowdown and, to get a unit determinant and match \cref{eq:vafaaut},
\begin{equation}
\label{eq:outautd4twist}
    w \to - w.
\end{equation}. The corresponding generators of the induced action on $H_{2}(Y_{D_4},\mathbb Z)$ are
\begin{equation}
    M_1 =\left(
\begin{array}{cccc}
 0 & 1 & 0 & 0 \\
 1 & 0 & 0 & 0 \\
 0 & 0 & 1 & 0 \\
 0 & 0 & 0 & 1 \\
\end{array}\right), \quad
M_2 = \left(
\begin{array}{cccc}
 1 & 0 & 0 & 0 \\
 0 & 0 & 1 & 0 \\
 0 & 1 & 0 & 0 \\
 0 & 0 & 0 & 1 \\
\end{array}
\right), \quad
M_3 = \left(
\begin{array}{cccc}
 0 & 0 & 1 & 0 \\
 0 & 1 & 0 & 0 \\
 1 & 0 & 0 & 0 \\
 0 & 0 & 0 & 1 \\
\end{array}
\right).
\end{equation}
It is convenient, to determine the geometric counterpart of outer automorphism of the generic $D_j$, to concentrate on a certain $\mathbb{Z}_2 \hookrightarrow S_{3}$. We can take, without generality loss, the $\mathbb{Z}_2$ that exchanges $s_1 \leftrightarrow t_1$ in \cref{eq:d4singular}, and introduce the following coordinates: 
\begin{equation}
    z = \frac{s+t}{2^{2/3}}, \qquad y = 
   \frac{s-t}{2^{2/3} i}, 
\end{equation}
in these coordinates, \cref{eq:d4singular} reads
\begin{equation}
    \label{eq:d4singconvenient}
    x^2 - z y^2 - z^3 = 0,
\end{equation}
and the $\mathbb Z_2$ biholomorphism reads $x \to -x, y \to -y$ (as in \cref{eq:vafaaut}).
\subsubsection{\texorpdfstring{$D_{j+1}$}{} singularities}
To get the outer automorphism formula for the generic $D_{j+1}$ singularity, we can proceed by induction. Let us consider the $D_{j+1}$ equation: 
\begin{equation}
\label{eq:djplus1sing}
    x^2 - z y^2   + z^j = 0,
\end{equation}
we can first blow-up, with the toric ambient space formalism, the origin. The partial resolution of the singularity reads
\begin{equation}
    \label{eq:partialresdj}
    x_1^2 - y_1^2 z_1 \delta_1 + z_1^j \delta_1^{-2+j} = 0
\end{equation}
with toric weights $[x_1] = [y_1] = [z_1] = 1, [\delta_1] = -1$ and the SR condition reading $(x_1,y_1,z_1)\neq \vec{0}$; the blowdown map is
\begin{equation}
    \label{eq:partialblowdownmapdj}
    (x_1,y_1,z_1,\delta_1) \longrightarrow (x,y,z)  = (x_1 \delta_1,y_1 \delta_1,z_1 \delta_1).
\end{equation}
We can see that, if we set $z_{1}$ equal to some non-zero constant (that in particular amounts to pass to the patch $z_1 \neq 0$), \cref{eq:partialresdj} describes a $D_{j-1}$ divisor; if we set $y_1$ to a constant, we see, at leading order, an $A_{1}$ singularity\footnote{One might be tempted to set also $\delta_{1}$ to constant value, the equation apparently has the shape (in the $x_{1}, y_1,z_1$ variables) of a $D_{j+1}$ singularity, reached when $x_{1}, y_1,z_1$ are all equal to zero. However, this point is excluded by the SR condition.}. Stated differently, the blowup of the origin of a $D_{j+1}$ partially resolves the singularity as
\begin{equation}
    D_{j+1} \to D_{j-1} \oplus A_{1}
\end{equation}
The outer automorphism of the $D_{j+1}$ is hence induced by the outer automorphism of the $D_{j-1}$ and, on the homology two-cycles, acts as 
\begin{equation}
\label{eq:homologyactiondj}
    [\mathbb P^1_1] \leftrightarrow [\mathbb P^1_{2}],
\end{equation}
where we are following the notation of \cref{tab:DynkinADE}.

If $j+1$ is even, the outer automorphism reads $y_{1} \to - y_{1}$, if it is odd, $x_{1} \to - x_1$. Inserting it into the blowdown map, \cref{eq:partialblowdownmapdj}, we see that the outer automorphism reflects $x \to -x$ for $j+1$ odd, and $y \to -y$ for $j+1$ even.  We can further "twist" the actions, in order to get \cref{eq:vafaaut}, sending also $x_1 \to -x_{1}$ (resp. $y_1 \to -y_1$) for the $j+1$ odd (resp. even) case. This further twist is just required to match \cref{eq:vafaaut}, getting a transformation with determinant one on the ambient space coordinates, but does not change the homology action, \cref{eq:homologyactiondj}.
\subsubsection{\texorpdfstring{$E_6$}{} singularity}
Finally, we consider $Y_{E_6}$, the last ADE singularity whose Dynkin diagram enjoys a $\mathbb Z_2$ outer automorphism reflecting the nodes with center the trivalent node 
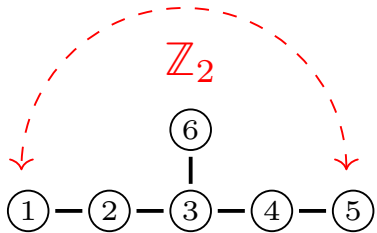
\begin{figure}[H]
\centering
    \scalebox{1.8}{\begin{tikzpicture}
        \draw[dashed,red,<->] (-1.05,4.1) arc (0:180:1.2cm);
        \node at (-2.2,4.9) {\footnotesize {$\textcolor{red}{\mathbb Z_2}$}};
        \draw (-3.4,3.8) circle (0.15);
        \draw[thick] (-3.2,3.8)--(-3.0,3.8);
        \draw (-2.8,3.8) circle (0.15);
        \draw[thick] (-2.6,3.8)--(-2.4,3.8);
        \draw (-2.2,3.8) circle (0.15);
        \draw[thick] (-2,3.8)--(-1.8,3.8);
        \draw (-1.6,3.8) circle (0.15);
        \draw[thick] (-1.4,3.8)--(-1.2,3.8);
        \draw (-1,3.8) circle (0.15);
        \draw[thick] (-2.2,4.0)--(-2.2,4.2);
        \draw (-2.2,4.4) circle (0.15);
        \node at (-3.4,3.8) {\tiny$1$};
        \node at (-2.8,3.8) {\tiny$2$};
        \node at (-2.2,3.8) {\tiny$3$};
        \node at (-1.6,3.8) {\tiny$4$};
        \node at (-1,3.8){\tiny$5$};
        \node at (-2.2,4.4){\tiny$6$};
        \end{tikzpicture}
        }
        \caption{$E_6$ Dynkin diagram with its outer automorphism}
        \label{fig:e6dynkin}
        \end{figure}

.\\ \indent  The  $E_6$ singularity is defined as 
\begin{equation}
\label{eq:e6singularity}
x^2 - y^3 - z^4 = 0.
\end{equation}
Its full resolution is the following hypersurface in toric ambient space \cite{Collinucci:2020jqd}:
\begin{equation}
\label{eq:e6resolved}
w^2 - v_1 v_3 v_2^2 s^3 + v_1^2 t^4  \qquad \subset \qquad 
\renewcommand{\arraystretch}{1.3}
\begin{array}{cccccccc}
\hline
 s_1 & t_1 & w & v_1 & v_2 & v_3 &  v_4\\
\hline
 1 & 1 & 1 & -1 & 0 & 0 & 0   \\
 1 & 0 & 1 & 1 & -1 & 0 & 0   \\
  0 & 0 & 1 & 1 & 1 & -1 & 0   \\
   0 & 0 & 1 & 1 & 0 & 1 & -1  \\
\end{array}
\end{equation}
whose SR ideal is 
\begin{equation}
    \label{eq:sre6}
    (s,t,w) \neq 0, \text{ and }  (s,w,v_1) \neq 0 \text{ and } (w,v_1,v_2) \neq 0 \text{ and } (w,v_1,v_3)  \neq 0.
\end{equation}
The blowdown map is 
\begin{equation}
    x = w v_1 v_2^2 v_3^4 v_4^6, \quad y = s v_1 v_2^2 v_3^3 v_4^4, \quad z = t v_1 v_2 v_3^2 v_4^3.
\end{equation}
The exceptional $\mathbb P^1s$, labeled as in \cref{fig:e6dynkin}, are
\begin{eqnarray}
\label{eq:curvese6}
&&\mathcal C_{1} \equiv \left\{v_2 = w + v_1 t^2 =
 0\right\} \nonumber \\
&& \mathcal C_{2} \equiv \left\{v_3 = w + v_1 t^2 =
 0\right\} \nonumber \\
&& \mathcal C_{3} \equiv \left\{v_4 = w^2 - v_1 v_3 v_2^2 s^3 - v_1^2 t^4 =
 0\right\} \nonumber \\
&& \mathcal C_{4} \equiv \left\{v_3 = w - v_1 t^2 =  
 0\right\} \nonumber \\
 && \mathcal C_{5} \equiv \left\{v_2 = w - v_1 t^2  =  
 0\right\} \nonumber \\
 && \mathcal C_{6} \equiv \left\{v_1 = w =  
 0\right\} \nonumber \\
\end{eqnarray}
One can see that the action
\begin{equation}
    \label{eq:automorphisme6}
    \left\{w \to - w, t \to -t\right\} \Rightarrow \left\{x \to -x,z \to -z\right\}
\end{equation}
induces the correct automorphism, \cref{eq:vafaaut}, of the Dynkin diagram. One can also note that the reflection $t \to - t$ is not necessary to produce the desired exchange of the $\mathbb P^1$s, but is needed if we want the biholomorphism to have unit determinant.

\subsubsection{Action on the relative homology}
\label{sec:actionrelativehom}
We conclude this section considering the action of the biholomorphism on the relative cycles. We concentrate on the $A_j$ singularity (following closely the discussion in \cite{Witten:2009xu}), the other Du Val singularities can be studied analogously. Furthermore, the case of $G_2$ manifold can also be studied with the results presented in this section, taking, as relative cycles, the direct product between the relative cycles of the Du Val and the $\mathbb S^3$ that is the basis of the Du Val fibration.

In the $A_j$ case \cite{Witten:2009xu}, the relative homology is generated by $j+1$ non-compact two-cycles $C_i$ defined as follows: given the deformed $A_j$, 
\begin{equation}
\label{eq:relativecycles}
    x y = z^{j+1} + u_2 z^{j-2} + ... u_{2j+1} = (z - z_1) ... (z - z_{j+1}), 
\end{equation}
$C_i$ is a circle fibration over a real path that starts from $z_i$ and goes to infinity in the $\mathbb C \ni z$-plane. The precise shape of the path is not important, as long as it intersects just one of the $z_i$. The $C_i$ intersect each other as\footnote{Note that this basis is not the (usual) dual basis of the compact homology in an algebraic sense, as we have $C_i \cdot C_{j,j+1} = \delta_{i,j} - \delta_{i,j+1}$.}
\begin{equation}
    C_i \cdot C_k = \delta_{ik}.
\end{equation}Any resolution of an $A_j$ singularity can be rewritten, with an hyperk\"ahler transformation, as \cref{eq:relativecycles} with all the $z_i$ lying on (say) the real axis. Then, the transformation in \cref{eq:vafaaut} acts as\footnote{We can understand the minus sign in \cref{eq:transflaw} as follows: an outgoing path from $z_i$ will still be an outgoing path (from $z_{j+2-1}$). However, the the exchange $u \leftrightarrow v$ in \cref{eq:vafaaut} flips the orientation of the non-compact cycles. The minus sign in \cref{eq:transflaw} is consistent with the fact that instead the compact cycles preserve their orientation under \cref{eq:vafaaut}.} 
\begin{equation}
\label{eq:transflaw}
    C_i \leftrightarrow -C_{j+2-i}.
\end{equation}
Furthermore, we can write the compact cycles in terms of the $C_i$ as 
\begin{equation}
\label{eq:embeddinghom}
    C_{i,k} \equiv  [C_{i}] - [C_k], \quad [\mathbb P^1_i]  = C_{i+1,i},
\end{equation}
where the $C_{i+1,i}$, with $i = 1,...,j$, generate $H_2(Y_\mathfrak g,\mathbb Z)$.
The compact cycles $C_{i,k}$ satisfy, by  definition, the following condition
\begin{equation}
\label{eq:homrelaux}
    C_{i,l} + C_{l, k} + C_{k,i} = 0.
\end{equation}
For non-compact topological spaces, regular homology is dual to relative homology via Lefschetz duality, thus
\begin{equation}
    \label{eq:noncompacthomasdual}
    H_k(Y_{A_j},\partial Y_{A_j}) \cong H^{n-k}(Y_{A_j}) \cong \text{Hom}\Big(H_{n-k}(Y_{A_j},\mathbb Z),\mathbb Z\Big),
\end{equation}
where in the case at hand, $n=4$ and $k=2$.\footnote{We also use the fact that $H_*(Y_{\mathfrak{g}})$ is torsion free for the last congruence.} 
The dual space is the space of linear maps  that assign to each compact cycle an integer, i.e. $f(C_{i,j})=b_{i,j} \in \mathbb{Z}$. From  \cref{eq:homrelaux}, it follows by applying $f$ that the coefficients must be of the form $b_{i,j}=b_i - b_j$. Therefore, the set of maps $f$ is labeled by sequences of integers $\{b_i | i=1,...,j+1\}$ up to shifts of an arbitrary integer\footnote{Indeed, a $j+1$-tuple $b_i$ and $b_i + b$ act in the same way on the compact homology}
\begin{equation}
\label{eq:equivalencerel}
    b_i \sim b_i + b.
\end{equation}
The function $f$ can be expressed in terms of the cycles $C_i$: 
\begin{equation}
\label{eq:expansion}
    f := C = \sum_{i=1}^{j+1} b_i C_i,
\end{equation}
acting on the compact cycles via the intersection pairing: 
\begin{equation}
    C' \to C\cdot C' = f(C'), \qquad C' \in H_2(Y_{A_j},\mathbb Z). 
\end{equation}

Summing up, we have  
\begin{equation}
    H_{2}(Y_{A_j},\partial Y_{A_j}) = \frac{\langle C_1,...,C_{j+1}\rangle_{\mathbb Z}}{\sim} = \frac{\mathbb Z^{j+1}}{\sim},
\end{equation}
    and
\begin{equation}
    H_{2}(Y_{A_j},\mathbb Z) = \langle C_{1,2},...,C_{j,j+1}\rangle_{\mathbb Z} \subset H_{2}(Y_{A_j},\partial Y_{A_j}),
\end{equation}
with the equivalence relation given by \cref{eq:equivalencerel}. The relative homology group can be identified with the weight lattice of the $A_{j}$ Lie algebra, while the regular one with the root lattice. By the geometric engineering dictionary, these data can be used to study extended operators. In particular, given the action in \cref{eq:transflaw} on non-compact cycles, we see the explicit action of the isometry symmetry on Wilson and 't Hooft operator, obtained by wrapping branes on the cycles $C_i$.\footnote{Let us note that, among the cycles in $H_{2}(Y_{A_j},\partial Y_{A_j})$, there are also cycles associated to endable operators, e.g. Wilson lines that can be screened by gluons. These lines are the ones that belong to the kernel of the quotient that defines the defect group.}

Finally, let us see how the action of \cref{eq:vafaaut} acts on the topological operators of the defect group via \cref{eq:transflaw}. We consider
\begin{equation}
\label{eq:relhom}
    \Big(H_{2}(Y_{A_j},\partial Y_{A_j})/H_{2}(Y_{A_j},\mathbb Z)\Big).
\end{equation}
To visualize \cref{eq:relhom}, it is useful to reason in terms of the expansion in \cref{eq:expansion}. $H_{2}(Y_{A_j},\mathbb Z)$ embeds in $H_{2}(Y_{A_j},\partial Y_{A_j})$ via \cref{eq:embeddinghom}, and all the $C_{i,j} = C_i - C_{j}$ are compact cycles represented by tuples of the form $[(0,..., 1,...,-1,...)]$ (all zero entries, except for the $i$-th and the $j$-th ones). Consequently, we can always rewrite, by adding terms in the compact homology, the equivalence class of a $j+1$-tuple $[(b_1,...,b_{j+1})]$ in \cref{eq:relhom} as $[(B,...,0)]$, with $B = \sum_i b_i$. Then, we can use \cref{eq:equivalencerel} to identify 
\begin{equation}
    B \sim B + (j + 1), 
\end{equation}
in agreement with
\begin{equation}
\label{eq:finalhomology}
\mathbb Z_{j+1} = H_1(\partial Y_{A_j},\mathbb Z) = \Big(H_{2}(Y_{A_j},\partial Y_{A_j})/H_{2}(Y_{A_j},\mathbb Z)\Big).
\end{equation}
We can now understand the effect of \cref{eq:vafaaut} on the topological defects implemented by branes. Calling $\epsilon$ the generator of \cref{eq:relhom},  we have 
\begin{equation}
    \epsilon = [(1,...,0)]\to - [(0,...,1)] = -[(1,...,0)] = -\epsilon = j \epsilon,    
\end{equation}
where the arrow is the transformation law of \cref{eq:transflaw}. Hence, a defect brane  $D_{\epsilon}$ wrapped over $\epsilon$ transforms as  $D_\epsilon \to D_{-\epsilon} = D_{j \epsilon}$ under \cref{eq:vafaaut}, making \cref{eq:vafaaut} a good candidate for the geometric realization of charge-conjugation.
\subsection{General remarks on 0-form symmetries as isometries}
\label{sec:autgroupsurfaces}
In the first part of this section, we will comment on the possible mixing of the biholomorphisms that we listed so far with the "quasi-homogeneous biholomorphisms" of \cref{eq:quasihomgeneral} of $X$, interpreting it as a mixing of the discrete 0-form symmetry with R-symmetries. In the second part we will comment on the relation between biholomorphisms, isometries and 0-form symmetries in the context of geometric engineering.

In the previous sections, we found good biholomorphisms to represent the outer automorphism symmetries in theories engineered with geometries built out of Du Val singularities. For $A_{2j-1}, D_n, E_6$ we picked the selected biholomorphisms asking the determinant of the transformation on the ambient space coordinates to be 1. One might wonder if, in general, other choices are available, and how to list all of them. There are infinitely many holomorphic volume forms associated to the same complex structure (parametrized by a $\mathbb C^*$ rescaling); consequently, requiring the biholomorphism to have unit determinant is too restrictive.

In order to list all the possible biholomorphisms of a Du Val (or, in general, of an hypersurface) singularity $P(x) = 0$, we look for transformations $x \to x'$ that respect
\begin{equation}
    \label{eq:autlist}
    P(x'(x)) = \lambda P(x),
\end{equation}
with $\lambda \in \mathbb C$ a constant. We can restrict to the case 
\begin{equation}
    \label{eq:coordtransfold}
    x'_j(x) = M_{j}^k x_k,\qquad M_j^k \in Gl(3,\mathbb C[[x]]),
\end{equation}
where $\mathbb C[[x]] \equiv \mathbb C[[x_1,...,x_N]]$ denotes the formal power series\footnote{We can exclude Laurent power series, as they would send some points to infinity, and hence can not be biholomorphisms of the hypersurface.} with center the origin of the ambient space $\mathbb C^N$ (for Du Val singularities, $N = 3$). Indeed, we used the non-compact shift part of the biholomorphisms group of the ambient space to fix the isolated singular point to be at the origin of the ambient space. Furthermore, we are using that all the biholomorphisms of an hypersurface can be described in terms of coordinates of the ambient space (and are hence induced by the restriction of an analytic transformation of it). We would like now to get rid of the infinite series; to achieve so, we use that the singularity is quasi-homogeneous, namely it is preserved by the action in \cref{eq:quasihomgeneral}. In particular, a biholomorphism has to commute with the quasi-homogeneous action, hence we can not have higher order terms
\begin{equation}
    x_1 \to x_1' =  x_{1} + c_1 x_{1}^2 + ...
\end{equation}
as $x_1$ and $x_1^2$ do not have the same quasi-homogeneous weight (and so on). We can now have two cases: 
\begin{enumerate}
    \item "non-resonant" case: for each $x_j$, there is no $x_k$ such that $w_{x_j} = w_{x_k} n$, with $n \in \mathbb N$; 
    \item "resonant" case: otherwise.
\end{enumerate}
In the non-resonant case, we then have that $M$ is a \textit{diagonal} matrix with constant coefficients: 
\begin{equation}
\label{eq:coefficientsrestr}
    M \in Gl(3,\mathbb C).
\end{equation}
In the resonant case, we can have mixing between $x_j$ and terms like $x_k^n$. As an example, in the case of Du Val singularities of type $A_n$, the coordinates $x,y$ has equal weight, and hence can mix (but $x$ (similarly $y$) can not mix with $x^m$ or $y^m$, for $m> 1$), hence we can again take $M$ to satisfy \cref{eq:coefficientsrestr}, but in this case (as it is indeed for \cref{eq:vafaaut}), $M$ is non-diagonal.\footnote{One can rule out explicitly other resonant cases, e.g. in the $E_6$ case $x$ and $z^2$ have the same quasi-homogeneous weight, but this does not produce new non-trivial biholomorphisms.}
Summing up, in the special case of Du Val singularities, all the biholomorphisms take the form: 
\begin{equation}
    \label{eq:coordtransf}
    x'_j(x) = M_{j}^k x_k,\qquad M_j^k \in Gl(3,\mathbb C),
\end{equation}
In order to pass to the special unitary biholomorphisms, we simply impose det$(M) = 1$. 
In the case of Du Val singularities, we explicitly checked that
\begin{itemize}
    \item each of the biholomorphisms with unit determinant coincide with one of the automorphisms of the Dynkin diagram;
    \item the det$(M) \neq 1$ biholomorphisms can be always written as a composition of a quasi-homogeneous transformation and a $\text{det}(M)=1$ one;
    \item the induced action on the first homology is insensitive to the quasi-homogeneous part of the transformation, as this will simply lift to the resolved Du Val singularity as $e_i \to \beta^{[e_i]} e_i $, with $[e_i]$ the induced weight on the toric ambient space. In particular, the aforementioned action does not permute the loci $e_i = 0$, and hence acts trivially on the exceptional cycles.
\end{itemize}
We can then conclude that for the Du Val singularities the biholomorphisms with determinant different from one can be understood as a composition of one of the biholomorphisms of the previous sections and a quasi-homogeneous rescaling. The induced action on the two-cycles is not affected by the choice of the quasi-homogeneous transformation.

We can make sense, physically, to the apparent ambiguity of choosing the biholomorphisms with unit-determinant, by characterizing physically the quasi-homogeneous transformations. To do so, we move away from the origin of the metric moduli space of the Du Val, by performing either a complex deformation or a resolution. Considering, e.g., the $A_j$ singularity (the other Du Val can be treated analogously), we concentrate on its complex deformations: 
\begin{equation}
\label{eq:deformationAnaux}
    x^2 + y^2 + z^{j+1} + u_2 z^{j-1} + ... + u_{j+1} = 0.
\end{equation}
The weights of the quasi-homogeneous transformations in \cref{eq:quasihomgeneral} are
\begin{equation}
    w_x = \frac{1}{2}, \quad w_y = \frac{1}{2}, \quad w_{z} = \frac{1}{n+1}. 
\end{equation}
After turning on the deformation parameters $u_{k}$ in \cref{eq:deformationAnaux}, the quasi-homogeneous action is not anymore a biholomorphism of the deformed Du Val, and "spurionically" modifies the deformation parameters as\footnote{For complex deformations, the coordinates on the moduli space of vacua are (invariants) polynomials of the periods of the holomorphic volume form. For quasi-homogeneous biholomorphisms, the transformation on $u_j$ does not come from a reshuffling of the cycles, but rather from a rescaling of the holomorphic volume form. On the other hand, in the resolved phase, the differential form describing the geometry is the K\"ahler form, that, being real, is not rescaled by the $\abs{\beta} = 1$ quasi-homogeneous actions. As, concerning the quasi-homogeneous part, this is the case for all the biholomorphisms in \cref{eq:vafaaut}, we did not discuss the transformation of the K\"ahler form in the previous sections.} 
\begin{equation}
    u_{k} \to \beta^{k/(j+1)} u_{k}. 
\end{equation}
As we will see in \cref{sec:complexdefs}, the transformations with $\abs{\beta} = 1$ acts on the deformation parameters (equivalently, on the moduli space of vacua of the physical theory) as a $U(1)$ subgroup of the R-symmetry.\footnote{We thank I\~{n}aki Garc\'ia Etxebarria for explaining this point to us.} Hence, we can conclude that different choices of biholomorphisms  are related by an R-symmetry transformation of the geometrically engineered theory.
As the blowup of a Du Val can be re-interpreted, via an hyperk\"ahler rotation, as a \cite{Witten:1997kz} complex deformation,\footnote{We note that, on the contrary, a complex deformation can not always be interpreted as a blowup, even for Du Val singularities.} this statement holds true also in the \textit{resolved} phase. 
Finally, we notice that, in principle, the method we presented allows us to compute the biholomorphisms group of isolated quasi-homogeneous n-fold singularity. We postpone a more precise statement of this fact, in the threefold case,  for further future investigation.

Let us conclude with a remark on the relation between the biholomorphisms and the isometries of the Ricci-flat metric of $X$. In fact, until now we realized the discrete zero-form symmetries as biholomorphisms of $X$. A natural question is then whether these biholomorphisms are also isometries of its CY metric.
 First, in the context of geometric engineering, the physically relevant quantity is the metric $g_{ij}$,  and  biholomorphisms and isometries of $X$ are not coincident sets.\footnote{By this we mean that it is not given that an isometry is a biholomorphism (or vice versa).} More precisely, if $X$ is K\"ahler, we have
\begin{equation}
    \label{eq:kahlmetric}
    g_{ij} = \omega_{ik}I^{k}_j,
\end{equation}
with $\omega$ and $I$ being, respectively, the K\"ahler form and the complex structure tensor.\footnote{This complex tensor is an endomorphism of the tangent bundle, squaring to minus the identity, and should not be confused with the holomorphic volume form.} From \cref{eq:kahlmetric}, we see that we might have transformations
\begin{equation}
    \label{eq:freedomaut}
    \omega \to R^{T} \omega R, \quad I \to R^{-1} I R,
\end{equation}
such that $R$ preserves the metric, but does not preserve the pair $(\omega,I)$: not all the isometries are biholomorphisms. Similarly, not all the biholomorphisms are isometries, as they could, in general, exchange holomorphic subvarieties with different K\"ahler volumes. For example, in the previous sections we considered biholomorphisms of the Du Val surfaces that permute the exceptional $\mathbb P^{1}$s; if two of them have different K\"ahler volumes, then the biholomorphism is not a isometry, and we interpret it as a broken symmetry. In the language of \cref{eq:freedomaut}, such biholomorphisms preserve $I$ but not $\omega$. However, the problem simplifies in the full singular phase\footnote{In general, we will have non-trivial subvarieties of the full moduli space of vacua, where the discrete zero-form symmetry is unbroken.}, where the holomorphic and K\"ahler volumes of the compact cycles of the internal geometry are sent to zero (equivalently, we sit at the origin of the moduli space of vacua of the geometrically-engineered theory). 
The most general biholomorphism will always induce a linear action on the homology of the deformed/resolved CY, and hence it fixes the origin of the moduli space. Consequently, if some kind of uniqueness theorem applies for the metric also for non-compact CYs, then the biholomorphism is also an isometry of the most singular phase of $X$: the pair $(\omega,I)$ (equivalently,  $\omega$ and holomorphic volume form\footnote{This second pair contains the same information of $(\omega,I)$, up to identifying $\Omega$ with $\gamma \Omega$, with $\gamma \in \mathbb C^*$.} $\Omega$) is preserved by the biholomorphism, and so it is the \textit{unique} Ricci-flat metric. This uniqueness property holds true for all the Du Val singularities with ALE metric (see \cite{Chen:2019snz} for a review of recent results in this sense), that is the one most naturally used in the geometric engineering setup.\footnote{Another possibility is the ALF metric (know also as the Taub-NUT metric), also in this case the uniqueness result is guaranteed at least in the $A_n$ and $D_n$ cases (see \cite{Chen:2019snz} and references therein).} This result ensures that all the biholomorphisms, at least in the context of geometric engineering with Du Val and with the $G_2$ manifolds that we consider in this work, are also isometries of the singular metric. For what concerns general CY threefolds instead, the uniqueness of the Ricci-flat metric given the cohomology classes of the K\"ahler form and of the holomorphic volume form is not guaranteed, even after specifying boundary conditions at infinity on the metric (see e.g. \cite{szekelyhidi2020uniqueness} for a discussion on this). We anticipate here that, however, a uniqueness result for the Ricci-flat metric holds also for the \textit{toric} Gorenstein CY$_3$ that we study in \cref{sec:toricCY3}. Hence, also in that case, biholomorphisms of the most singular phase of the CY$_3$ are isometries of the corresponding (unique) Ricci-flat metric. 

 Finally, we notice that the biholomorphisms that we are considering, and in particular the associated\footnote{From now to the end of this section, we assume of being at the origin of the moduli space of the associated theory.} isometries, induce isometries of the link $\mathbf{L}_X$ of the considered singularities. This can be proved as follows. When, as in \cref{eq:conedefinition}, the metric is conical we can have two kind of isometries for $X = \text{Cone}(\mathbf{L}_X)$: those rescaling the radial coordinate $r \to \lambda r$, with $\lambda \in \mathbb R_{+}$, and the isometries inherited from those of $\mathbf{L}_X$ (with $\lambda = 1$). The latter act fixing $r$ and transforming $\mathbf{L}_X$. Let us denote with $\Psi$ a biholomorphism associated with a certain finite 0-form symmetry, then\footnote{In the case of $D_4$ singularity, the statement still holds true, as \cref{eq:cyclicity} holds for each of the biholomorphisms associated with the $S_3$ 0-form symmetry group.} 
\begin{equation}
\label{eq:cyclicity}
    \Psi^k = \mathbb{1}
\end{equation}
for some $k \in \mathbb N$. This automatically tells us that $\Psi$ can not act on the radial coordinate of the cone, as the action $r \to \lambda r \to \lambda^2 r \to ...$, being $\lambda$ a real positive number different from one, is clearly not cyclic. This statement holds for all the biholomorphisms considered in this paper, and is also true in general if the 0-form symmetry group is finite. We can sum up the previous discussion as follows: 
\begin{enumerate}
    \item In general, biholomorphisms might not be isometries and viceversa.   
    \item However, the biholomorphisms considered in this work are isometries of the most singular phase of  $X$.
    \item Given $X$ that, as in this work, admits a unique conical Ricci-flat metric, then the isometry associated with a 0-form symmetry always preserves $\mathbf{L}_X$, acting isometrically on it.
\end{enumerate}
\section{Charge-conjugation for 7d and 4d SYM from geometry}
\label{sec:mixedanomalies7dsym}
In this section, we will see how the biholomorphisms of \cref{sec:resduval} reflect into the physical theories that can be geometrically engineered starting from internal manifolds built using Du Val singularities. In \cref{sec:7dsym} we discuss 7d SYM. Then, in \cref{sec:4dnequal1} we will study M-theory on a specific class of $G_2$ seven-manifold, engineering 4d $\mathcal N = 1$ SYM theories. Finally, in \cref{sec:symtft}, we will comment on the SymTFT viewpoint on the 0-form symmetries associated with the considered isometries. 
 
\subsection{7d SYM}
\label{sec:7dsym}
Let us consider M-theory on a Du Val singularity $Y_{\mathfrak g}$; the theory in the seven external dimension transverse to the Du Val is 7d $\mathcal N =1$ SYM, with gauge algebra $\mathfrak g$. In this section, we will first recap the construction of \cite{Albertini:2020mdx,Bhardwaj:2020phs}, that identifies the generator of 1-form symmetry with a specific compact divisor of the Geometric Engineering space. Then, by studying the action of the isometries on this divisor we will realize the geometric counterpart of the 2-group structure of the corresponding theories.  In this setup, the M-theory geometric engineering dictionary (on the CB of the 7d SYM) works as follows: 
\begin{enumerate}
    \item The reduction of the M-theory three-form $C_{3}$ over the normalizable $\gamma_{i} \in H^2(Y_{\mathfrak g},\mathbb Z)$ are the photons in the Cartan subalgebra $\mathfrak h \subset \mathfrak g$. 
    
    \item The volumes of the exceptional cycles are given by the integral of the three K\"ahler form of the $Y_{\mathfrak g}$ over the $\gamma_i$. These can be interpreted as the vevs of the scalars of the 7d vector multiplet parametrizing the Coulomb branch.
    
    \item The set of elements of $H_{2}(Y_{\mathfrak g},\mathbb Z)$ that support a BPS M2-brane is isomorphic to the root system of $\mathfrak g$. By this we mean that an element $\nu = \sum_{i=1}^{\text{rank}(\mathfrak g)} \nu_{i} \gamma_i \in H_{2}(Y_{\mathfrak g},\mathbb Z)$ (with $\nu_i \in \mathbb Z$) wrapped by an M2 brane, engineers a BPS state (namely, a W-boson of the 7d theory), if and only if $(\nu_1,...,\nu_{\text{rank}(\mathfrak g)})$ is a root of $\mathfrak g$. Furthermore, the gauge charges of the $\nu$ W-boson are exactly given by $(\nu_1,...,\nu_{\text{rank}(\mathfrak g)})$.
\end{enumerate}
We are interested in studying the interaction between the $\mathbb Z_2$-isometry and the 1-form symmetries of the theory. To do so, we first  study the geometric counterpart of the 1-form symmetry generator with the following procedure \cite{Albertini:2020mdx}: 
\begin{enumerate}
    \item Find the Cartan element $\mathcal Z$ of the 7d theory that corresponds to $\mathcal Z(G)$, with $G$ the simply connected group with Lie algebra $\mathfrak g$. 
    
    \item This Cartan element will be a \textit{integer-coefficients} combination of the generators $h_{i} \in \mathfrak h$ associated, with the geometric-engineering dictionary, to the $\gamma_i$. We can hence write, keeping in mind that the geometric engineering dictionary must be applied,
    \begin{equation}
        \mathcal Z = \sum n_i \gamma_i, 
    \end{equation}
    with $n_i \in \mathbb Z$.
    
    \item Finally, we can use the Poincar\'e duality between the $n_{i}$ and the $H_{2}(Y_{\mathfrak g},\mathbb Z)$, to construct a divisor 
    \begin{equation}
        D_{\mathcal Z} = \sum_{i} n_i S_i, 
    \end{equation}
    where $S_i$ are the codimension-two cycles Poincar\'e duals to the normalizable two-forms $\gamma_i$. As discussed in \cref{sec:bulk-boundarycycles}, this compact divisor can be associated to the torsional cycle generating the 1-form symmetry. The action of the isometry on $D_{\mathcal Z}$ can therefore be interpreted, in the field theory context, as the action of a 0-form symmetry on the 1-form symmetry one.
 
\end{enumerate}
The computation \cite{Bhardwaj:2020phs} of the coefficients $n_j$ gives the following results:
\begin{itemize}
    \item in the $A_j$ case, the $D_{\mathcal Z}$ reads
\begin{equation}
    \label{eq:oneformdivaj}
    D_{\mathcal Z} = \sum_{i=1}^j i S_{i},
\end{equation}
where the biholomorphism associated with the outer automorphism exchanges $S_i \leftrightarrow S_{j+1-i}$. 
    \item in the $D_{2j}$ case, we have $\mathcal Z(SO(4j)) \cong \mathbb Z_2 \times \mathbb Z_2$. The divisor corresponding to the generators of the two $\mathbb Z_2$ factors are
\begin{equation}
    \label{eq:oneformdivdj}
    D_{\mathcal Z,1} = \sum_{i=1}^{2j-1} \frac{(1 - (-1)^i)}{2} S_{i}, \quad 
    D_{\mathcal Z,2} = S_{2j} + \sum_{i=1}^{2j-2} \frac{(1 - (-1)^i)}{2} S_{i}, 
\end{equation}
and the outer automorphism exchanges $D_{\mathcal Z,1} \leftrightarrow D_{\mathcal Z,2}$.
    \item in the $D_{2j+1}$ case, the center of $SO(4j+2)$ is $\mathbb Z_4$ and the divisor corresponding to its generator reads
\begin{equation}
\label{eq:oneformdivd2nplus1}
    D_{\mathcal Z} = 3 S_{2j+1} + S_{2j} + \sum_{i=1}^{2j-1} (1 - (-1)^i) S_{i},
\end{equation}
We note that the role of the two $A_1$ tails of the $D_{2j+1}$ Dynkin diagram is arbitrary and, as expected by the biholomorphism described in \cref{sec:resduval}, can be permuted in \cref{eq:oneformdivd2nplus1}.

\item For the $E_6$ case, the center of the simply-connected Lie group is $\mathbb Z_3$, and the divisor associated to its generator is 
\begin{equation}
\label{eq:oneformdive6}
D_{\mathcal Z} = \sum_{i=1}^5 i S_i.
\end{equation}
\end{itemize}

For the case of $Y_{\mathfrak g}$, we identify the $S_{i}$ with the $\mathbb P^1$s, numbered as in \cref{tab:DynkinADE}, that resolve the Du Val singularities.

We can now give a physical interpretation of the zero-form symmetry associated to the isometry presented in \cref{sec:resduval}. The charge-conjugation of 7d $A_{j}$ SYM acts on the adjoint fields as 
\begin{equation}
\label{eq:gaugethminustransposedtransf}
    \Phi_{i} \to - g^{-1} \Phi_{i}^T g, \qquad i = 1,2,3, 
\end{equation}
where $g \in SU(j+1)$ is an inner automorphism. The choice of this inner automorphism is arbitrary, and its action is defined up to Weyl reflections. Let us consider for example the $A_2$ case, we can pick 
\begin{equation}
\label{eq:convenientformA2}
 g =    \left(
\begin{array}{ccc}
 0 & 0 & 1 \\
 0 & -1 & 0 \\
 1 & 0 & 0 \\
\end{array}
\right).
\end{equation}
Since the homology $H_{2}(Y_{\mathfrak g},\mathbb R)$ is identified with the Cartan subalgebra, to check that indeed \cref{eq:gaugethminustransposedtransf} and \cref{eq:convenientformA2}  match the geometric isometry action of \cref{sec:resduval}, we can study how they act on the Cartan subalgebra $\mathfrak h \subset A_{2}$, spanned by\footnote{We note that the the sum in the Lie algebra does not correspond to the sum in the homology. Instead, the sum (with integer coefficients) in the homology means "adding the roots" and moves us inside the root system. In particular, geometry is not sensible to the coefficients along a given root-eigenspace of $\mathfrak g$.} 
\begin{equation}
    h_{1} = \left(
\begin{array}{ccc}
 1 & 0 & 0 \\
 0 & -1 & 0 \\
 0 & 0 & 0 \\
\end{array}
\right) \quad h_{2} = \left(
\begin{array}{ccc}
 0 & 0 & 0 \\
 0 & 1 & 0 \\
 0 & 0 & -1 \\
\end{array}
\right).
\end{equation}
It is easy to see that \cref{eq:gaugethminustransposedtransf}, with $g$ as in \cref{eq:convenientformA2}, exchange $h_{1} \leftrightarrow h_2$, and acts on the roots $\alpha_{1,2}$ as 
\begin{equation}
    \alpha_1 \leftrightarrow \alpha_2 \, .
\end{equation}
In order to guess the convenient form we used the fact that, if we want to order the fundamental representation $\textbf{3}$ from the highest to the lowest weight, after we sent $\Phi \to - \Phi^T$ we have to reshuffle the basis elements of the fundamental representation (because what used to be the highest weight is now the lowest, and so on). This also tells us how to generalize \cref{eq:convenientformA2}: let us consider the following elements of $A_{j}$: 
\begin{equation}
    \label{eq:vkdefinition}
    \left(v_{k}\right)_{a,b} = \delta_{a,k} \delta_{b,k+1}, \qquad \left(\widetilde{v}_{k}\right)_{a,b} = \delta_{a,k+1} \delta_{b,k},
\end{equation}
with $a,b,= 1,..,j+1$ the indeces of the fundamental representation of $A_j$, and $k=1,..., j$. Each $v_k,$ (resp. $\widetilde{v}_k$) is the nilpositive (resp. nilnegative) element of $A_j$  of the standard $\mathfrak{su}(2)$ triple associated with the $k$-th root.\footnote{The standard triple associated to the $k$-th root of the $A_j$ Lie algebra is, in \cite{CollingwoodMcGovern1993} conventions, the $\mathfrak su(2)$ subalgebra of $A_j$ formed by $v_k, \widetilde{v}_k$, and their commutator $H_k = [v_k,\widetilde{v}_k]$. The $v_k,\widetilde{v}_k$, taken together, generate $A_j$ as a Lie algebra.}  To generalize \eqref{eq:convenientformA2}, we then look for a matrix $g$ such that 
\begin{equation}
    g^{-1} v_{k} g = v_{(j+1)-k}, \qquad g^{-1} \widetilde{v}_{k} g = \widetilde{v}_{(j+1)-k},
\end{equation} and, when $j$ is even,\footnote{We note that all of this is just "kinematic", as we simply choose a convenient generator of the outer automorphism to match the geometric expectation.} 
\begin{equation}
    g^{-1} v_{j/2} g = -v_{j/2}, \qquad g^{-1} \widetilde{v}_{j/2} g = -\widetilde{v}_{j/2}.
\end{equation}
A similar result can be generalized also to the $D_{j}$ and $E_6$ cases \cite{Trautner:2016ezn}: the outer automorphism of the Dynkin diagram is, up to an internal automorphism, associated with the charge-conjugation of the 7d theory. 

As a concluding remark, let us stress that these isometry symmetries are not given by branes wrapped on torsional cycles. The case of $SU(n)$ SYM makes this explicit: the boundary geometry is given by the lens space $S^3/\mathbb{Z}_n$, which has only $n$-torsional 1-cycles. Branes wrapped on these cycles lead to $\mathbb{Z}_n$ symmetries, while the automorphism of interest is always a $\mathbb{Z}_2$ symmetry of the field theory, even for odd $n$. Moreover, the dimension of branes in M-theory is not compatible with the dimension of the support of a 0-form symmetry in 7d, preventing us from realizing a construction similar to the one of \cite{Dierigl:2023jdp}. This is but an example of an extension of the “symmetry/torsional cycle" paradigm presented in \cite{GarciaEtxebarria:2022vzq ,Heckman:2022muc} that can help to further deepen the link between geometry and field theory. 

\subsection{4d \texorpdfstring{$\mathcal N =1$}{} SYM}
\label{sec:4dnequal1}
The discussion of the previous section translates almost untouched to the 4d $\mathcal N = 1$ SYM theory with gauge algebra $\mathfrak g$. The theory can be geometrically engineered via M-theory on the $G_2$-manifold $M_{7,\mathfrak g}$, that is a fibration of $Y_{\mathfrak g}$ on the $\mathbb S^3$ \cite{Acharya:2000gb,Acharya:2001hq,Atiyah:2000zz}:
\begin{equation}
\begin{tikzcd}
Y_{\mathfrak g}  \arrow[r, hook] & M_{7} \arrow[d,two heads] \\
& \mathbb S^3
\end{tikzcd}
\end{equation}
The boundary geometry contains the information about the higher form symmetry of the theory. From \cite{Albertini:2020mdx}, we have that $H_*(\partial M_{7}, \mathbb{Z})=(\mathbb{Z},\Gamma,0,\mathbb{Z},\Gamma,0,\mathbb{Z})$, where $\Gamma$ is the center of the algebra $\mathfrak{g}$. We can conclude, as expected from the field theory analysis, that theory admits two 1-form symmetry mutually non-local, given by M2 wrapped on torsional 1-cycles and M5 wrapped on torsional 4-cycles.

From the analysis of the previous section, we know that a biholomorphism of the Du-Val surface becomes an isometry for specific resolutions and, in this case, it acts trivially on the $\mathbb S_3$ base of the fibration. Let us note, however, that in this case, as expected from the fact the field theory is a 4d $\mathcal{N}=1$ theory, there is no modulus associated to the resolution of the Du Val singularity. It turns out that the singularity is partially smoothed out by a non-zero volume for the base $\mathbb S_3$ corresponding to the fact that in the IR the field theory confines. The volume of the base becomes proportional to the gaugino VEV \cite{Atiyah:2000zz}, an hallmark of confinement. This means that the isometry is never broken by any choice of VEV, since the Du Val singularity remains singular and therefore the symmetry associated to the isometry is preserved along the RG-flow.

Consequently, the same analysis of \cref{sec:7dsym} carries through and the zero-form symmetry action can be interpreted as the 4d charge-conjugation.

Let us note that, as in \cite{Kaidi:2021xfk, Bhardwaj:2022yxj, GarciaEtxebarria:2022vzq,Antinucci:2022eat,Arias-Tamargo:2022nlf}, that upon gauging the 0-form symmetry, a non-intrinsic non-invertible symmetry can be realized. We leave the study of the geometric counterpart of this non-invertible symmetry to future work.

\subsection{SymTFT in the presence of isometries}
\label{sec:symtft}
As we mentioned in the previous sections, the non-trivial action of the isometries on the exceptional cycles can be interpreted as an higher group structure between a 0- and a 1-form symmetry on the engineered field theory. Before delving into the example, let us remark that the GE construction provides only a “relative" realization of the field theory \cite{Freed:2006yc, GarciaEtxebarria:2019caf, Albertini:2020mdx}. A global form is then decided by fixing the boundary condition for the operators implementing the different symmetries \cite{Albertini:2020mdx}. Thus one should bear in mind that when we speak about a gauge theory, we are referring to the corresponding {\it algebra} rather than to the corresponding {\it group}, until boundary conditions are fixed.

For concreteness, and to make contact with previous literature \cite{Bhardwaj:2022yxj, GarciaEtxebarria:2022vzq}, let us consider the case of $\mathcal{N}=1$ $Spin(4n)$ SYM. This theory can be engineered by following the construction described in \cref{sec:4dnequal1} with a D$_{2n}$-type Du Val singularity fibered over a 3-sphere, where we fix the global variant to be the simply connected one. As we discussed previously, there is an isometry of the Du Val singularity acting non-trivially on the generators of $\mathbb{Z}_2 \times \mathbb{Z}_2$ 1-form symmetry. This is captured in the SymTFT that schematically reads
\begin{align}\label{eq:anom_lagr}
    \mathcal{L}_{\text{SymTFT}}= \tilde B_2 \dd B_2 + \tilde C_2 \dd C_2 + \tilde A_3 \dd A_1 + p A_1 \wedge B_2 \wedge C_2 \, ,
\end{align}
where $A_1$ is the background for the 0-form symmetry, while $B_2$ and $C_2$ are backgrounds for the 1-form symmetry. The coefficient $p$ is an integer defined mod 2, that is non-zero if there is a non trivial mixing of the operators.\footnote{The periodicity of $p$ is determined by methods described in \cite{Wang:2014pma}. Even if we cannot provide the exact value of such coefficient, from our analysis we know that it cannot be 0 mod $2$ when the 0-form acts non-trivially on the 1-form symmetry.}

One can then follow the analysis of \cite{Bhardwaj:2022yxj, GarciaEtxebarria:2022vzq} in order to reach all different realizations of the various global variants of this theory. Indeed, all such variants are connected by (non-anomalous) discrete gauging of the different background fields. For example, the action of the 0-form symmetry on the 1-form one leads to a 2-group symmetry when the global variant for the gauge group is $Spin(4n)$, while we have a mixed anomaly for the $SO(4n)$ one and possibly non-invertible symmetries for other global variants.\footnote{A similar reasoning can be applied to the $SU(n)$ and $E_6$ case as discussed in \cite{Bhardwaj:2022yxj, Arias-Tamargo:2022nlf , Bhardwaj:2023kri}.}

As remarked in the introduction, \cref{eq:anom_lagr} is telling us that there is a non trivial interplay between 0- and 1-form symmetries. Let us note that when the 0-form symmetry is absent, the SymTFT is a BF theory that does enjoy automorphism that can be interpreted as symmetries of the topological theory itself. These automorphisms are associated to dualities of the underlying dynamical theory \cite{Kaidi:2022cpf}, that are enhanced to actual symmetries for certain choices of moduli. This has the effect of changing the “free" BF theory to an interacting one, enriching the structure of possible boundary conditions. Therefore, in analogy with \cite{Kaidi:2022cpf}, the Lagrangian in \cref{eq:anom_lagr} can be interpreted as a theory described by twisted cocycles. It is therefore tempting to identify the object that encodes the SymTFT in GE with not only $H_*(\mathbf{L}_X)$, but with $H_*(\mathbf{L}_X)_{\rho}$, i.e. the boundary homology twisted by the action of isometries. Despite it would be interesting to make this statement more explicit, in this paper we limit ourselves to pointing out this possibility given our analysis of isometries and their action on torsional cycles. The relation between twisted homology, dualities and diffeomorphism is left for future works.

Before ending this section, we make a final comment about the relation of boundary conditions between geometry and field theory. While the boundary conditions analysis is solid in the context of field theory, the geometrical counterpart remains unexplored. Indeed, while going from one boundary condition to another amounts in a series of discrete gauging on the field theory side, this is less understood in the geometrical picture. As anticipated, the operation of discrete gauging might have non-trivial consequences on the geometry. For example in \cite{Argyres:2016yzz,Arias-Tamargo:2019jyh, Arias-Tamargo:2023duo}, it is shown that upon discrete gauging some $\mathcal{N}=2$ theories, the flavor symmetry might change. In GE terms, this means that the gauging procedure might lead to a change in the structure of codimension-one singular loci. We leave the exploration of these phenomena to future work.

\section{Higher structure of 5d SCFTs from toric CY3}
\label{sec:toricCY3}
The previous analysis can be easily generalized to the case of 5d SCFT engineered via M-theory on toric CY$_3$, which are the focus of this section.

In the toric CY$_3$  case, despite the equations defining the manifold are usually difficult to deal with, one can use graphical methods to study crepant resolutions. Indeed, different resolutions are in one-to-one correspondence with triangulations of the toric diagram encoding the geometry. In particular, points inside the polygon correspond to compact divisors, 4-cycles, while internal edges to compact curves, 2-cycles, \cite{Aharony:1997bh, Jefferson:2017ahm, Jefferson:2018irk, Closset:2019juk}.

\begin{figure}
    \centering
    \includegraphics[scale=0.4]{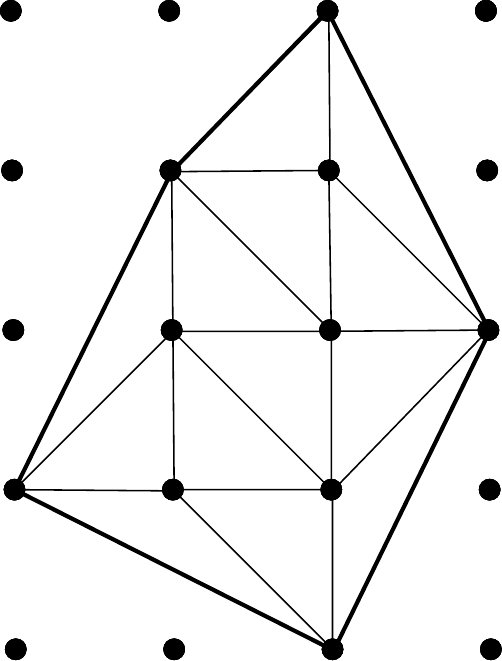}
    \caption{A triangulated toric diagram.}\label{fig:toric_generic}
\end{figure}

As in the previous section, one can consider a divisor $D_{\mathcal Z}$ associated to the boundary cycles encoding the 1-form symmetry. In this case, a $\mathbb{Z}_2$ isometry of the manifold, associated with charge conjugation, can be seen directly from the toric diagram. Indeed, if via an $SL(2,\mathbb{Z})$ transformation the toric diagram can be put in a shape that is invariant under the biholomorphism $(x,y) \to (-x,-y)$, then the theory is invariant under charge conjugation\footnote{In fact, this transformation corresponds, in the dual (p,q) web, to applying twice the $S$-transformation. This operation is the IIB realization of the charge-conjugation symmetry.} \cite{Closset:2018bjz}. The biholomorphism exchanges the compact divisors in a similar fashion to the one described in the previous sections. Again, one has to be careful about the fact that the biholomorphisms in general do not coincide with isometries. However, also in the case of \textit{Gorenstein} toric\footnote{In the context of geometric engineering, it is usually assumed that the CY3 is Gorenstein. Other options, associated to the so-called $q$-Gorenstein case, are not common at all in the literature, and their study goes beyond the scope of this paper.} CY$_3$, there exists a unique toric Ricci-flat metric for fixed K\"ahler class and complex structure \cite{cho2008uniqueness}. Hence, analoguesly to the Du Val and to the $G_2$-manifold case that we studied, we can claim that biholomorphisms become isometries after we blow-down all the compact cycles of the threefold.  

From this action we can thus infer that there is an action of charge conjugation on the 1-form symmetry. This fact can be seen explicitly in the case of 5d $SU(p)_q$ singularities engineered via $Y^{(p,q)}$ threefolds. Following \cite{Albertini:2020mdx}, the singularities are described by the following toric diagram
\begin{figure}
    \centering
    \includegraphics[scale=0.4]{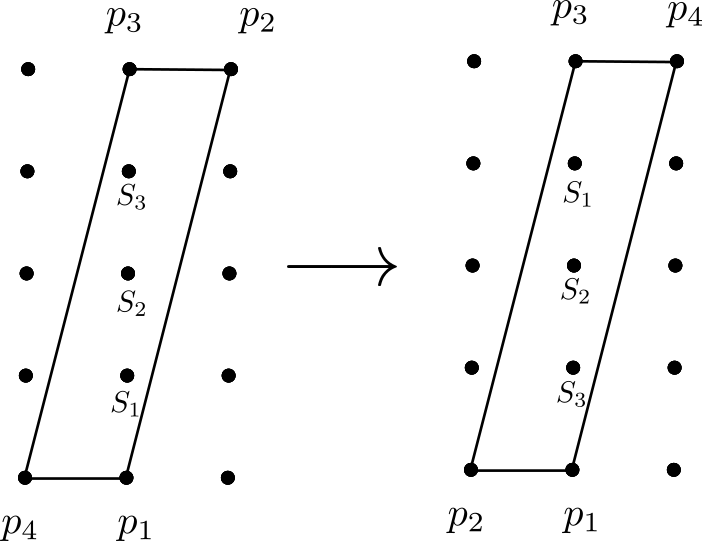}
    \caption{Toric diagram corresponding to 5d $SU(n)_0$ and its mapping under charge conjugation.}\label{fig:toric_generic1}
\end{figure}
\begin{align}
    p_1=(0,0) \, , \quad p_2=(1,p-q) \, , \quad p_3=(0,p) \, , \quad p_4=(-1,0) \, .
\end{align}

The compact divisors, $S_i$, have coordinates $(0,i)$, $i=1,...,p-1$ and  under the $\mathbb{Z}_2$ biholomorphism, the divisors are mapped as $S_i \leftrightarrow S_{p-1-i}$. This action is analogous to the one described in the previous section on ADE singularities. Indeed, as shown in \cite{Albertini:2020mdx}, the divisor $D_{\mathcal{Z}}$ associated to the 1-form symmetry is given by $D_{\mathcal{Z}} = \sum_i S_i$. 

For 5d SCFTs that do not admit a Lagrangian phase, it is still possible to define the divisor $D_{\mathcal Z}$. To do so, one needs to consider a generic (integral) linear combination of the compact divisors of the toric CY$_3$ such that all the compact curves have intersection number multiple of $n$\footnote{The computation of these intersection numbers can be carried out algorithmically in with Sage, we thank Azeem Hasan for help in the writing of the code.}, the degree of the 1-form symmetry, i.e. all dynamical BPS particles have a charge multiple of $n$ with respect to a diagonal $U(1)$ factor on the Coulomb branch.

\begin{figure}
    \centering
    \includegraphics[scale=0.4]{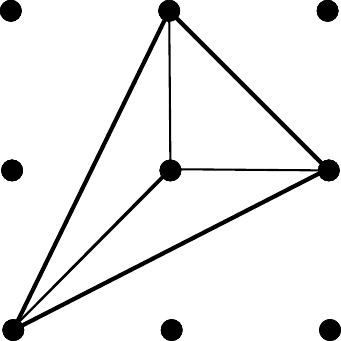}
    \caption{The toric diagram of the Seiberg $E_0$ theory, it admits a triality symmetry with trivial action on the 1-form symmetry.}\label{fig:toric_generic2}
\end{figure}

As an example, let us consider M-theory on the "local $\mathbb P^2$ threefold\footnote{This threefold is the total space of the anticanonical bundle over $
\mathbb P^2$.},  the simplest non-Lagrangian 5d SCFT, also known as the Seiberg $E_0$ theory. This theory is engineered via the local $\mathbb{P}^2$ singularity described by the toric diagram in \cref{fig:toric_generic2}
and the intersection matrix of curves (up to equivalences) and divisors is given by
\begin{center}
\begin{tabular}{c|c|c|c|c}
     curve / divisor  & $D_1$ & $D_2$ & $D_3$ & $S_0$ \\
     \hline
     C & 1 & 1 & 1 & -3
\end{tabular} \, .
\end{center}
In this case, it is trivial to see that $D_{\mathcal Z} = S_0$.

\subsection{Particle-instanton symmetry}

A feature of 5d SCFTs engineered via toric geometry is that the $SL(2,\mathbb{Z})$ duality group of the theory is manifest in the toric diagram: two theories are dual if their toric diagrams can be deformed one onto the other via an $SL(2,\mathbb{Z})$ transformation. In particular, as discussed previously, a duality may become a symmetry if the transformed toric diagram is identical to the starting one. The case of charge conjugation is the simplest instance, but in principle one can have geometries invariant under any finite subgroup of $SL(2,\mathbb{Z})$, i.e. $\mathbb{Z}_{2,3,4,6}$. These subgroups are generated respectively by $S^2, ST^{-1}, S$ and $ST$, and their effect on the field theory is to exchange particles and instantons.

\begin{figure}
    \centering
    \includegraphics[scale=0.4]{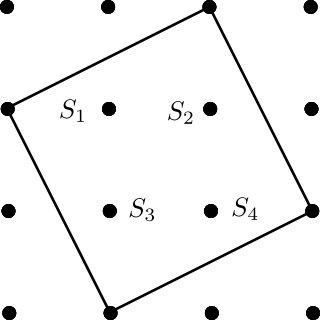}
    \caption{Example of a non-Lagrangian 5d theory that admits a tetrality symmetry with non-trivial action on the 1-form symmetry.}\label{fig:toric_generic3}
\end{figure}

As before, we can expect a mixing between the n-ality symmetry, i.e. the $\mathbb{Z}_n$ particle-instanton symmetry, and the 1-form symmetry. This can be checked by computing the divisor $D_{\mathcal Z}$ and see what is the n-ality action on it. The $E_0$ theory is already an example, despite being a trivial one, since the triality symmetry acts trivially on $S_0$. As a non trivial example let us consider the toric singularity in \cref{fig:toric_generic3}.

This theory has a tetrality symmetry, rotating the diagram by 90 degrees. The center symmetry is generated by $D_{\mathcal Z} = S_1 + 2 S_2 + 3 S_3 + 4 S_4$ and both charge conjugation and the tetrality symmetry acts in a non trivial way, respectively by exchanging $S_1\leftrightarrow S_4 \, , \quad S_2\leftrightarrow S_3$ and $1 \to 2 \to 4 \to 3 \to 1$. The action of $S$ exchanges particles and instantons, exchanging F1 strings and D1 branes stretching in the pq-web \cite{Aharony:1997ju}. We will not delve into further examples, since it boils down to building toric diagrams symmetric under the action of different subgroups of $SL(2,\mathbb{Z})$, but we end this section noticing that our analysis allows one to find how particles-instantons symmetries that arise in certain 5d field theories can be captured by isometries that in turns have non-trivial actions on the higher form symmetries of the theory. In particular, being purely geometrical, our analysis works also for non-Lagrangian theories.

\section{Complex deformations and 0-form symmetries}
\label{sec:complexdefs}
In the previous sections, we saw how the biholomorphisms of Du Val singularities act on the cycles of the resolved surface. However, already considering the 7d SYM case, we have an R-symmetry $\mathfrak{so}_R(3)$ that rotates among each other the three real adjoint scalars appearing in the $\mathcal{N}=1$ multiplet. As we will briefly review in \cref{sec:duvalasspectral}, the Casimir invariants of these adjoint scalars control the K\"ahler volumes of the resolved $\mathbb{P}_1$s w.r.t. to the three inequivalent K\"ahler forms associated to the hyperk\"ahler structure of the Du Val \cite{Katz:1992aa,Collinucci:2022rii}. In this sense, since the singularity is hyperk\"ahler, one can perform a rotation of the complex structures, passing from the \textit{resolved} phase to the \textit{deformed} one.\footnote{The complex structure can be described by picking two out of the three K\"ahler forms, and considering the periods of a complex-valued combination of them.} It is thus natural to study the action of isometries also on the deformed phase of Du Val singularities. In the following, we will explore the action of biholomorphisms proposed in the previous sections on the full moduli space of two different kind of theories: the 7d SYM theories in \cref{sec:SYMfinal}, and the 6d (2,0) SCFT in \cref{sec:6d20full}, focusing on the deformed phase of the considered geometries.

\subsection{Du Val as spectral varieties}
\label{sec:duvalasspectral}
Let us quickly recap the technology of \cite{Katz:1992aa,Collinucci:2022rii} for describing deformed Du Val singularities. Let us focus on the $A_j$ case, with the $D,E$ cases being straightforward generalizations.  

A generic (versal) deformation of an $A_{j}$ singularity can be written as 
\begin{equation}
\label{eq:spectralvarietyAn}
    x^2 + y^2   = z^{j+1} + u_2 z^{j-1}+...+u_j,
\end{equation}
with $u_2,...,u_{j+1} \in \mathbb C$ the (versal) deformation  coefficients. This deformed geometry admits a description as a spectral variety, namely as a variety described by the eigenvalues of a certain matrix. Given an $A_{j}$-algebra valued complex matrix $\Phi$, we can compute the following 
\begin{equation}
    \label{eq:branelocusascharacteristicpoly}
    x^2 + y^2  = \text{Det}(z \mathbb{1} - \Phi),
\end{equation}
and we can match the above expression with \cref{eq:spectralvarietyAn} giving the following identification \cite{Katz:1992aa}
\begin{equation}
    \label{eq:casimirtodefs}
    u_k = \frac{\text{Tr}(\Phi^k)}{k}+...
\end{equation}
where the dots denote higher powers of appropriate lower-degree Casimir invariants, $\text{Tr}(\Phi^{J})^K$,  such that $J + K = k$. In other words, the Casimir invariants of $\Phi$ control the deformation of the $A_{j}$ singularity, \cref{eq:spectralvarietyAn}.

In mathematics, the complex structure of a deformed Du Val can also be described via the periods of the holomorphic\footnote{In this case, actually the "holomorphic-symplectic" volume form.} volume form $\omega = \omega_1 + i \omega_2$ on the $\mathbb S^2$s that \textit{deforms} the Du Val singularity.  These topologically two-spheres intersect each other as the Dynkin diagram of the corresponding Lie algebra, playing a role similar to the $\mathbb P^1$s resolving the Du Val singularity. However, differently from the $\mathbb P^1$s of the resolved phase, they are not holomorphically embedded in the deformed $Y_{\mathfrak g}$, and so have a non-zero holomorphic volume
\begin{equation}
\label{eq:holvolumes}
V_i \equiv \int_{\mathbb S^2_{i} 
\subset \widetilde{A_{j}}} \omega,
\end{equation}
with $i = 1,...,\text{rank}(H_2(Y_{\mathfrak g},\mathbb Z)) = \text{rank}(\mathfrak g)$.
These holomorphic volumes have an easy description in terms of $\Phi$, indeed, we have \cite{Katz:1992aa}
\begin{equation}
    \label{eq:holvolumes2}
    \alpha_{i}-\alpha_{i+1} = V_i,
\end{equation}
where $\alpha_i$ are the eigenvalues of $\Phi$. The ordering of the eigenvalues is arbitrary, but there is no puzzle related to this: reshuffling the eigenvalues amounts to picking a different basis of two-cycles of the deformed geometry, and is associated with Weyl reflections.

\subsection{7d SYM and the deformed Du Val singularities.}
\label{sec:SYMfinal}
We can now understand, in the familiar setup of 7d SYM theories, how the spectral geometric description of Du Val singularities finds a physical counterpart.

The scalar part of the 7d SYM multiplet consists of three adjoint scalar fields $\Phi_i$, $i = 1,2,3$ in the Lie algebra of the \textit{compact} Lie group\footnote{Stated differently, we are \textit{not} taking the complexification of the Lie algebra.}. As always, simultaneous diagonal VEVs of these fields parametrize the vacua of the theory, and the full moduli space is given by
\begin{equation}
\label{eq:modulispace7dsym}
    \mathcal M_{\mathfrak g}=\mathbb R^{3 r}/\mathcal W,
\end{equation}
with $r$ the rank of the 7d gauge algebra $\mathfrak g$ and $\mathcal W$ being its Weyl group, that acts, e.g. in the $A_j$ case, by permuting the eigenvalues of the $\Phi_i$.

Let's us first consider a vacuum with $\Phi_1 = \Phi_2 = 0$; we can describe the corresponding Du-Val geometry using \cref{eq:branelocusascharacteristicpoly}. When all the eigenvalues of $\Phi_3$ have the same phase, the moduli space is described by a \textit{deformed} singularity which can be rotated into a \textit{resolved} one by means of a hyper-K\"ahler rotation, see \cite{Witten:1997kz} for details.

Therefore, whenever we are in a \textit{deformed} phase which can be rotated into a \textit{resolved} one, we can use the analysis of the previous sections to recover the action of the biholomorphisms on the cycles. However, this is not true for a generic complex deformation, that thus cannot be interpreted as a blowup. These cases are described by \textit{generic} VEVs of the operator $\Phi = \Phi_1 + i \Phi_2$. For a generic point on the moduli space we can then study the effect of (now spontaneously broken) biholomorphisms. In this case, since we are interested in realizing the charge conjugation of the physical theory, its action on the complex scalar is $\Phi_i \to -g(\Phi_i)^T g^{-1}$. This induces an action of the deformation parameters of \cref{eq:casimirtodefs}, given by
\begin{equation}
u_k \to (-1)^ku_k,
\end{equation}
correspondingly modifying \cref{eq:branelocusascharacteristicpoly}. This action we just described can be matched with the ones described in \cref{sec:resduval}. Indeed, we have that the 2-cycles we are introducing by deforming the singularity are exchanged as 
\begin{equation}
    \label{eq:volumepermutation}
    V_{i} = \alpha_i - \alpha_{i+1} 
    \quad \longrightarrow \qquad \alpha_{n-i-1} - \alpha_{n-i} = V_{n-1-i}
\end{equation}
and for $A_{2l+1}$, $V_{l+1} \to V_{l+1}$, leading to the same picture we described in \cref{sec:resduval}.

We can explicitly see that the biholomorphism of \cref{sec:resduval} acts giving a minus sign to the coefficients of the $z^{2k+1}$ terms in \cref{eq:spectralvarietyAn}. Again, one might argue that the action of the biholomorphism we introduced is not the one naturally associated to the charge-conjugation (that is expected to reflect the homology classes of the Du Val resolved/deformed singularity). However, as shown in \cref{sec:7dsym}, these two actions are equivalent up to Weyl-group action.\\
We conclude by noticing that the description of Du Val as spectral varieties can be exploited, both in the context of M-theory \cite{Collinucci:2021ofd, DeMarco:2021try, Collinucci:2022rii,DeMarco:2022dgh,Collinucci:2021wty} and IIB \cite{Klemm:1996bj,Wang:2015mra} geometric engineering, to build lower-dimensional SCFTs. In these cases, one considers complex one-parameter families\footnote{See also \cite{Sangiovanni:2024nfz} for a recent application in the context of 3d $\mathcal N = 2$ theories, and \cite{Acharya:2024bnt} for a discussion on the CY3 metrics in this setup.} of (deformed) Du Val singularities, producing (respectively) 5d $\mathcal N = 1$ rank-zero SCFTs and class-S theories. It would be interesting, as a possible follow-up to study the role of isometries in these setups. 
\subsection{A comment on 6d (2,0)}
\label{sec:6d20full}
Du Val singularities also appear in the IIB engineering of (2,0) theories. Indeed, one can consider type IIB on $\mathbb R^{1,5} \times Y_{\mathfrak g}$, that is well know to engineer the 6d (2,0) theory of type $\mathfrak g$ \cite{Witten:1997kz}. The (2,0) multiplet contains five real-scalars, valued in $\mathfrak g$, whose eigenvalues can be associated to the periods of, respectively, the RR and NSNS potentials and of the three K\"ahler forms associated to the hyperk\"ahler structure. These real-valued zero modes are rotated into each other by an $\mathfrak{so}(5)$ R-symmetry. As in the 7d SYM case, the moduli space of vacua of the theory is parametrized by Weyl-invariant polynomial of the aforementioned periods: 
\begin{equation}
    \label{eq:modulispace2comma0}
        \mathcal M = \frac{\mathbb R^{5r}}{\mathcal W},
\end{equation}
with $\mathcal{W}$ the Weyl group of $\mathfrak g$.

Similarly to the 7d case, we can arbitrarily choose one of the adjoint scalars in the (2,0) to control the K\"ahler volume of the exceptional $\mathbb P^1$s, say $\Phi_1$. We can then pick $\Phi_2 + i \Phi_3$ to control the holomorphic volumes of the two-cycles describing the complex deformation, while we chose $\Phi_4$ and $\Phi_{5}$ to control the RR and NSNS moduli \cite{Aganagic:2015cta}.  \footnote{Please note that $\Phi_4$ and $\Phi_5$ do not have an immediate geometric interpretation; we leave a more refined understanding of the geometric counterpart of these moduli to future work.} This subdivision is completely arbitrary, and our choice is not invariant under the R-symmetry $\mathfrak{so}(5)$ group. Again, following \cref{sec:SYMfinal} we can infer the action of the 0-form symmetry associated with the biholomorphism on the K\"ahler and complex structure moduli. On top of that, using an $\mathfrak{so}(5)$ R-symmetry rotation, one in principle can understand the action of the zero-form symmetry on the non-geometric $B_{NSNS}$ and $B_{RR}$.

Let's conclude this section by matching our result with the one of \cite{Dierigl:2023jdp}. In that paper, the authors realised the topological operators of the charge-conjugation symmetry of the 6d (2,0) SCFTs using the recently introduced type IIB R7-brane. From their analysis, the existence of the R7 brane can be predicted as follows: 
\begin{enumerate}
    \item the (2,0) theory can be realized with IIB on $\mathbb R^{1,5} \times Y_{\mathfrak g}$; 
    \item the (2,0) theory displays a $\mathbb Z_2$ zero-form symmetry that can be interpreted as charge conjugation, and hence there \textit{must} exist in IIB a stringy object that engineers the topological operator that generates this symmetry. 
    \item Requiring the topological operators generating the global symmetry to be realized by branes wrapped on (submanifolds of) the link $\mathbb S^3/\Gamma$ of the Du Val singularity, the IIB object realizing the aforementioned topological operator must be a brane that behaves exactly as the R7 brane of \cite{Dierigl:2023jdp}. 
\end{enumerate}
Our setup based on the isometries of the geometric engineering internal space is different from the ``brane at infinity" paradigm; however, the fact that we rediscover the presence of a charge conjugation symmetry is a further check of our construction. Moreover, from our analysis we are able to extend the result of \cite{Dierigl:2023jdp} by arguing that said symmetry can have non trivial action on the higher form symmetry of the theory.

As per the previous examples, we have that the biholomorphism of \cref{eq:vafaaut} acts on the torsional cycles on the base of the engineering geometry. Since in the 6d (2,0) theory there is a 2-form symmetry, implemented by topological D3 branes wrapped on said torsional cycles, we expect that the biholomorphisms can have a non-trivial action on these generators. Such action is manifest once the theory is reduced on a torus, leading to 4d $\mathcal{N}=4$ SYM, that indeed admits a charge conjugation symmetry for any value of the coupling $\tau$. We do not explore further the context of 6d theories and their SymTFT, mostly because they are relative theory and thus they generically do not admit topological boundary conditions for their topological operator. We leave this investigation to future work. It would be interesting to further compare our approach, based on the isometries of the internal space, with the R7 proposal of \cite{Dierigl:2023jdp}.

\section{Acknowledgments}
We are particularly thankful to Michele Del Zotto for suggesting the question that motivated this paper and for collaboration in the early stages of the work. We would like to thank Guillermo Arias-Tamargo, I\~{n}aki Garc\'ia Etxebarria, Giovanni Galati, Azeem Hasan, Max H\"ubner, Carlo Scarpa, and Alessandro Tomasiello for insightful discussions. The research of M.D.M. is funded through an ARC advanced project, and further supported by IISN-Belgium (convention 4.4503.15). M.D.M. has also received funding from the European Research Council (ERC) under the European Union’s Horizon 2020 research and innovation program (grant agreement No. 851931) during the initial stage of the work. The work of S.N.M. is supported by the Simons Foundation (grant \#888984, Simons Collaboration on Global Categorical Symmetries) and by the European Research Council (ERC) under the European Union’s Horizon 2020 research and innovation program (grant agreement No. 851931)

\begin{appendix}
\section{Resolution of \texorpdfstring{$D_{5}$}{} singularity}
To produce an example where we just have a $\mathbb Z_2$ outer-automorphism, we consider the $D_5$ singularity: 
\begin{equation}\label{eq:d5sing}
    x^2 - z y^2 + z^4 = 0.
\end{equation}
The crepant resolution of the singularity is 
\begin{equation}
    \label{eq:d5resolved}
    D_{3} D_4 s_2 t_{2}^2 + w_4^4 + D_3^2 s_2^4 \delta_2^4 = 0, 
    \qquad \subset \qquad 
\begin{array}{cccccccc}
 [w_4] & [s_2] & [t_2] & [D_3] & [D_4] & [\delta_2] & [\delta_4] \\
 \hline
 1 & 1 & 1 & -1 & 0 & 0 & 0 \\
 1 & 1 & 0 & 1 & 0 & -1 & 0 \\
 1 & 0 & 1 & 1 & -1 & 0 & 0 \\
 1 & 0 & 0 & 1 & 1 & 0 & 0 \\
 1 & 0 & 0 & 1 & 1 & 0 & -1 \\
\end{array}
\end{equation}
The SR condition leads us to exclude the zero loci of the following ideals
\begin{eqnarray}
    \label{eq:srconditiond5}
    I_1 = \left(\delta _2 s_2,\delta _4 D_4
   t_2,\delta _2 \delta _4^2 D_4
   w_4\right), \quad I_2 = \left(s_2,\delta _4^2 D_4 w_4,\delta _4^2
   D_3 D_4\right), \nonumber \\
   I_3  = \left(\delta _4 w_4,\delta _4
   D_3,t_2\right), \quad I_4= \left(w_4,D_3,D_4\right)
\end{eqnarray}
The blowdown map is
\begin{equation}
    \label{eq:blowupmapd5}
    x = D_3 D_4^2 w_4 \delta_2^2 \delta_4^4, \quad y =  D_3 D_4^2 t_2 \delta_2 \delta_4^3, \quad z = D_3 D_4 s_2 \delta_2^2 \delta_4^2.
\end{equation}
The exceptional $\mathbb P^1$ are
\begin{eqnarray}
\label{eq:curvesd5}
&&\mathcal C_{1} \equiv \left\{D_{4} = w_4 - i D_3 s_2^2 \delta_2^2 =
 0\right\} \nonumber \\
&& \mathcal C_{2} \equiv \left\{D_{4} = w_4 + i D_3 s_2^2 \delta_2^2\right\} \nonumber \\
&& \mathcal C_{3} \equiv \left\{D_4 = \delta_4 =
 0\right\} \nonumber \\
&& \mathcal C_{4} \equiv \left\{D_4 = D_3= 
 0\right\} \nonumber \\
 && \mathcal C_{5} \equiv \left\{D_4 = \delta_2 =  
 0\right\} \nonumber \\
\end{eqnarray}
We are now looking to an automorphism that exchanges $\mathcal C_{1} \leftrightarrow \mathcal C_{2}$, we have two options: $w_{4} \to -w_{4}$, and $D_{3} \to - D_3$. In the first case, the automorphism on the singularity reads $x \to - x$, in the latter $(x,y,z) \to - (x,y,z)$. Let us pick the first option, by adding a further twist $\delta_2 \to - \delta_2$ that does not modify the action on the homology of the resolved singularity \eqref{eq:d5resolved}. The automorphism induced on the singularity simply reads 
\begin{equation}
    \label{eq:automorphismonsingd5}
    x \to -x, \qquad y \to - y,
\end{equation}
and has determinant one. As we commented upon in \cref{sec:autgroupsurfaces}, we have many options of automorphisms, with determinant different from one, that induce the same action on the homology of the resolved $D_5$, but all of them are linked by a quasi-homogeneous rescaling of the Du-Val equation.

\end{appendix}

\phantomsection
\addcontentsline{toc}{section}{References}
\bibliography{cc_bib}{}

\end{document}